\documentclass[aps,prb,twocolumn,groupedaddress,showpacs,floatfix,nofootinbib]{revtex4}
\usepackage{graphics}
\usepackage{epsfig}
\usepackage{subfigure}
\usepackage{multirow}

\newcommand{\angs}{\mathrm{\AA}}
\newcommand{\magneton}{\mu_\mathrm{B}}

\newcommand{\be}{\begin{equation}}
\newcommand{\ee}{\end{equation}}

\newcommand{\bea}{\begin{eqnarray}}
\newcommand{\eea}{\end{eqnarray}}

\newcommand{\eqn}[1]{Eq.~(\ref{#1})}

\newcommand{\reffig}[1]{Fig.~\ref{#1}} 
\newcommand{\reffigs}[2]{Figs.~\ref{#1} and~\ref{#2}}

\newcommand{\reftab}[1]{Table~\ref{#1}}
\newcommand{\reftabs}[2]{Tables~\ref{#1} and~\ref{#2}}

\newcommand{\sizefactor}{\Omega_\mathrm{SF}}
\newcommand{\eform}{E_\mathrm{f}}

\newcommand{\emig}{E_\mathrm{m}}
\newcommand{\ebind}{E_\mathrm{b}}
\newcommand{\sfsub}{\sizefactor(\mathrm{sub})}
\newcommand{\eformsub}{\eform(\mathrm{sub})}
\newcommand{\eformfreesub}{\eform^\mathrm{free}(\mathrm{sub})}

\begin{document}

\title{Transition metal solute interactions with point defects in
  austenitic iron from first principles}

\author{D.J.~Hepburn}
\email[Email: ]{dhepburn@ph.ed.ac.uk}
\author{E. MacLeod}
\author{G.J.~Ackland}
\email[Email: ]{gjackland@ed.ac.uk}
\affiliation{Institute for Condensed Matter and Complex Systems, School of Physics and SUPA, The University of Edinburgh, Mayfield Road, Edinburgh, EH9 3JZ, UK.} 
\date{\today}
\pacs{61.72.-y,61.82.Bg,71.15.Mb,75.50.Bb}

\begin{abstract}

We present a comprehensive set of first principles electronic
structure calculations to study the properties of substitutional
transition metal solutes and their interactions with point defects in
austenite (face-centred cubic Fe). Clear trends were observed in these
quantities across the transition metal series. Solute-defect
interactions were found to be strongly correlated to the solute size
factors in a manner consistent with local strain field
effects. Functional relationships were determined in a number of
cases, although for some the early and late transition metal solutes
displayed quite distinct behaviour. Strong correlations with results
in ferrite (body-centred cubic Fe) were observed throughout, showing
insensitivity to the underlying crystal structure in Fe. We confirmed
that oversized solutes act as strong traps for both vacancy and
self-interstitial defects and as nucleation sites for the development
of proto-voids and small self-interstitial loops. The consequential
reduction in defect mobility and net defect concentrations in the
matrix explains the experimental observation of reduced swelling and
radiation-induced segregation in austenitic steels doped with
oversized solutes. These results raise the possibility that oversized
solutes remaining dissolved in oxide dispersion-strengthened (ODS)
steels after manufacturing could contribute to the observed
radiation-damage resistance of these materials. Our analysis of
vacancy-mediated solute diffusion demonstrates that Ni and Co diffuse
more slowly than Fe, along with any vacancy flux produced under
irradiation below a critical temperature, which is $400\pm 50$ K for
Co and their concentrations should be enhanced at defect sinks. In
contrast, Cr and Cu diffuse more quickly than Fe, against a vacancy
flux and will be depleted at defect sinks. Oversized solutes early in
the transition metal series form highly-stable solute-centred
divacancy (SCD) defects with a nearest-neighbour vacancy. The
vacancy-mediated diffusion of these solutes is dominated by the
dissociation and reassociation of the SCDs, with a lower activation
energy than for self-diffusion, which has important implications for
the nucleation and growth of complex oxide nanoparticles containing
these solutes in ODS steels. Interstitial-mediated solute diffusion is
energetically disfavoured for all except the magnetic solutes, namely
Cr, Mn, Co and Ni. Given the central role that the solute size factor
plays in the results discussed in this work, we would expect them to
apply, more generally, to other solvent metals and to austenitic
stainless steel alloys in particular.

\end{abstract} 

\maketitle

\section{Introduction}

The addition of major and minor alloying elements to steels has been
an essential technique for improving, amongst others things, their
mechanical, thermal and chemical properties for a particular
application throughout the entire history of iron and steel
manufacturing, research and technological progress. In the nuclear
industry the push to make the next generation of nuclear fission
reactors and prospective fusion reactors as safe and efficient as
possible places significant design constraints on the structural
materials used to build them. In particular, these materials must be
able to withstand higher temperatures, radiation doses and more
chemically corrosive environments than previous reactor systems,
whilst maintaining their mechanical integrity over timescales of half
a century or more.

One of the holy grails in nuclear materials is the so called
self-healing material, which exhibits few, if any, of the usual
problems found in irradiated materials, such as embrittlement, void
formation and swelling, radiation-induced segregation (RIS),
irradiation-induced creep (IIC) and irradiation-assisted stress
corrosion cracking (IASCC). In the early nineties Kato {\it et
  al.}\cite{Kato1991,Kato1992} made a significant step in the right
direction when they showed that the addition of around 0.35 at.\% of
oversized transition metal (TM) solutes, such as Ti, V, Zr, Nb, Hf and
Ta, to 316L austenitic stainless steel significantly reduced swelling
by both prolonging the incubation period for void nucleation to higher
doses and suppressing void growth and decreased the RIS of Cr away
from and Ni towards grain boundaries usually seen under
irradiation. Similar observations were also made by Allen {\it et
  al.}\cite{Allen2005} upon adding Zr to Fe-18Cr-9.5Ni austenitic
steel. Furthermore, it was observed that these beneficial effects
increased in strength with the size-factor of the solute, that is, in
the order, Hf$>$Zr$>$Ta$>$Nb$>$Ti$>$V\cite{Kato1991,Kato1992}.

Point defect (and in particular vacancy) trapping at the oversized
solutes was suggested as the primary mechanism behind the
observations\cite{Kato1991,Kato1992}, leading to a decrease in defect
mobility and net point defect concentrations, either via enhanced
recombination or the formation of secondary defects in the
matrix. Stepanov {\it et al.}\cite{Stepanov2004} demonstrated that a
model based on the trapping of vacancies by oversized solutes was
capable of reproducing the simultaneous suppression of RIS and void
swelling observed experimentally. The primary aim of the current work
is to improve upon the theoretical understanding of the mechanisms
underpinning the experimental observations of Kato {\it et
  al.}\cite{Kato1991,Kato1992} using detailed first-principle
calculations of the atomic-level processes involved.

The incorporation of small oxide nanoparticles, such as Y$_2$O$_3$, is
another important technique to strengthen and improve the
radiation-damage resistance of both
ferritic\cite{Kishimoto2009,Hsiung2010,Brodrick2014} and
austenitic\cite{Oka2011,Oka2013,Xu2011,Zhou2012,Gopejenko} steels,
allowing them to be used at higher temperatures and radiation dose
rates than standard steels. Small quantities of oversized solutes,
such as Ti and Hf, are commonly used in the formation of these ODS
steels to control the size of the oxide nanoparticles. While it is
generally accepted that the mechanical alloying techniques used in the
production of these steels fully dissolves the atomic components of
the Y$_2$O$_3$ and minor alloying element powders into the Fe matrix,
the subsequent nucleation and formation of the oxide nanoparticles
during heat-treatment and annealing is not completely understood. The
possibility for isolated, oversized solutes to remain dissolved in the
Fe matrix and contribute to the radiation-damage resistance of ODS
steels is also worthy of further investigation. We investigate both of
these questions within this work.

To the best of our knowledge, no first-principles calculations have
been performed to investigate the general behaviour of TM solutes or
their interactions with point defects in austenite. This is directly
related to the extensive computational effort required to explicitly
model the paramagnetic state of
austenite\cite{OlssonB,Alling,Kormann12,Steneteg} and to the large
number of near-degenerate reference states capable of modelling
metastable austenite at zero Kelvin\cite{Klaver12}. Density functional
theory (DFT) has been used to investigate the properties of Y in
austenite\cite{Gopejenko}, as a first step to understanding Y$_2$O$_3$
nanoparticle formation in ODS steel. The non-magnetic (nm) state of
face-centred cubic (fcc) Fe was used to model paramagnetic austenite,
in contrast to our previous first-principles studies in
austenite\cite{Klaver12,Hepburn13,Ackland11}, where magnetic effects
were included explicitly. In this work we have followed a similar
methodology by using the face-centred tetragonal (fct),
anti-ferromagnetic double-layer (afmD) collinear-magnetic state of Fe
to model austenite\cite{Klaver12,Hepburn13,Ackland11}. We have
investigated the properties of TM solutes in this state using
first-principles DFT calculations, in a comparable manner to the work
of Olsson {\it et al.} in the body-centred cubic (bcc) ferromagnetic
(fm) Fe ground state\cite{Olsson10}. In particular, we focus on solute
interactions with point defects and investigate any general trends
across the TM series and possible correlations between these
interactions and solute size-factors.

In section \ref{compDetailsSection} we present the details of our
method of calculation. We then proceed to discuss TM solute properties
in the defect-free lattice (section \ref{defectFreeSection}) and their
interactions with vacancy and self-interstitial defects (in sections
\ref{vacancySection} and \ref{siSection}, respectively) before making
our conclusions. A direct and fruitful comparison with results in bcc
Fe\cite{Olsson10} is made throughout. The TM solute data is summarised
in Appendix \ref{AppB}.

\section{Computational Details}
\label{compDetailsSection}

The calculations have been performed using the plane wave DFT code,
VASP\cite{KresseHafner,KresseFurthmuller}, in the generalised gradient
approximation with exchange and correlation described by the
parametrisation of Perdew and Wang\cite{PW91} and spin interpolation
of the correlation potential provided by the improved
Vosko-Wilk-Nusair scheme\cite{vwn}. Projector augmented wave (PAW)
potentials\cite{Blochl,KresseJoubert} were used for all TM
elements. First order Methfessel and Paxton
smearing\cite{MethfesselPaxton} of the Fermi surface was used
throughout with a smearing width, $\sigma = 0.2$ eV. Spin-polarised
(collinear magnetic) calculations have been performed for all magnetic
materials with local magnetic moments determined within VASP by
integrating the spin density within spheres centred on the atoms. The
sphere radii are given in Appendix \ref{AppA}.

A set of high-precision calculations were performed to determine the
ground state crystallographic and magnetic structures for all the TM
elements. A detailed account, including a short review of the
significantly more complex structure of Mn, is given in Appendix
\ref{AppA}, where the results are summarised along with previous
results for the fct afmD and fcc nm states of Fe\cite{Klaver12}. The
calculated crystallographic parameters were found to be, typically,
within 1-2\% of the experimental values\cite{Kittel}. Elastic
constants for fct afmD Fe were calculated
previously\cite{Klaver12}. Using the same technique, we found that
those for fcc nm Fe are $C_{11}=423$ GPa, $C_{12}=217$ GPa,
$C_{44}=236$ GPa and the bulk modulus, $B=286$ GPa.

Supercells of 256 ($\pm 1$, $\pm 2$,...) atoms were used for the TM
solute calculations with supercell dimensions held fixed at their
equilibrium values and ionic positions free to relax. Single
configurations were relaxed until the force components were no more
than 0.01 eV/$\angs$. Nudged elastic band\cite{NEB98} (NEB)
calculations using a climbing image\cite{HenkelmanClimb00} and
improved tangent method\cite{HenkelmanTangent00} were also used to
determine migration barriers with a tolerance for energy convergence
of 1 meV. A $2^3$ k-point Monkhorst-Pack grid was used to sample the
Brillouin zone along with a plane wave cutoff energy of 350 eV in all
these calculations, which were found to allow formation, binding and
migration energies as well as inter-particle separations and local
moments to be determined accurately\cite{Klaver12}.

\begin{figure}[htbp]
\includegraphics[width=0.9\columnwidth]{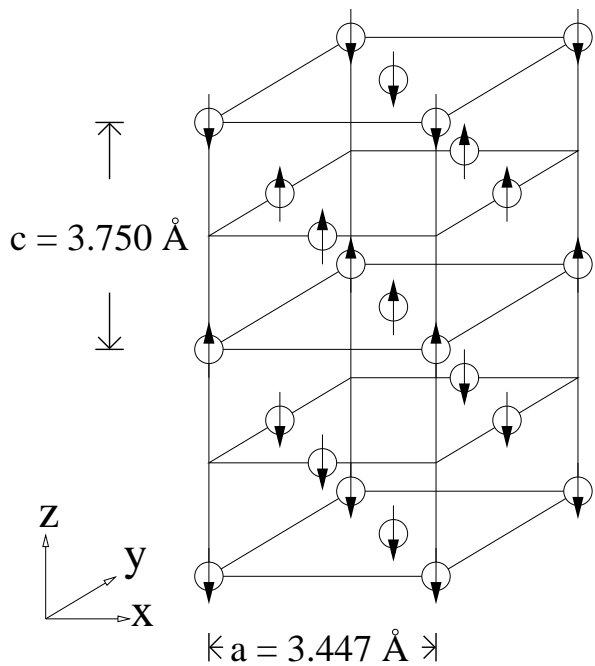}
\caption{\label{fctafmDFig}The fct afmD structure of Fe. The arrows
  indicate the local moments on the atoms, showing the magnetic planes
  and double-layer magnetic structure. Lattice parameters, $a$ and
  $c$, are also given.}
\end{figure}

We model austenite (at $T$=0K) using fct afmD Fe, which is the most
stable collinear magnetic reference state structure. This structure
consists of ferromagnetically aligned (001) fcc planes of atoms, which
we refer to as magnetic planes, with an up,up,down,down double-layer
ordering of moments on adjacent planes along the c-direction, as shown
in \reffig{fctafmDFig}.

An important part of this study is a comparison of results in fct afmD
Fe with those in bcc fm Fe using data from the work of Olsson {\it et
  al.}\cite{Olsson10}. We have performed additional calculations in
bcc fm Fe to provide data for the elements Sc, Zn, Y, Cd, Lu and Hg
not covered in that study. These calculations were performed in a 128
atom supercell with a greater plane wave cutoff energy of 350 eV, a
finer $4^3$ Monkhorst-Pack k-point grid and a near-identical lattice
parameter to the previous study\cite{Olsson10} [see Appendix
  \ref{AppA}]. A comparison of results for the elements Ti, Cu, Zr,
Ag, Hf and Au, between our method and Olsson {\it et
  al.}\cite{Olsson10}, showed that formation energies differed by no
more than a few hundredths of an eV, which is more than sufficient for
our purposes.

We define the formation energy, $\eform$, of a configuration
containing $n_\mathrm{X}$ atoms of each element, X, relative to a set
of reference states for each element using 
\be 
\label{EformEquation}
   \eform = E-\sum_\mathrm{X} n_\mathrm{X}E^\mathrm{ref}_\mathrm{X}, 
\ee 

where $E$ is the calculated energy of the configuration and
$E^\mathrm{ref}_\mathrm{X}$ is the reference state energy for element
X. We take the reference energies to be the energies per atom in the
ground-state crystal structures for all elements except Fe, where the
energy per atom in the solvent structure, that is in fct afmD or fcc
nm Fe, has been used.

We define the binding energy between a set of $n$ species, $\{A_i\}$,
where a species can be a defect, solute, clusters of defects and
solutes etc., using the indirect method as
\be 
   \ebind(A_1,...,A_n)= \sum_{i=1}^n \eform(A_i) - \eform(A_1,...,A_n) 
\ee 
where $\eform(A_i)$ is the formation energy for the single species,
$A_i$, and $\eform(A_1,...,A_n)$ is the formation energy for a
configuration where the species are interacting. An attractive
interaction, therefore, corresponds to a positive binding energy. One
intuitive consequence of this definition is that the binding energy of
a species, $B$, to an already existing cluster (or complex) of
species, $\{A_1,\dots,A_n\}$, which we collectively call $C$, is given
by the simple formula,
\be
   \ebind(B,C) = \ebind(B,A_1,\dots,A_n) - \ebind(A_1,\dots,A_n).  
\ee
This result will be particularly useful when we consider the
additional binding of a vacancy or solute to an already existing
vacancy-solute complex.

The size factor, $\sizefactor$, for a substitutional solute, X,
in an alloy can be defined\cite{Straalsund1974} as the change in
volume, $\Delta V$, upon replacing an average alloy atom with an X
atom, expressed as a fraction of the average atomic volume per lattice
site, $V_\mathrm{ave.}$. Practically, it can be defined in terms of
the (partial) atomic volume of solute X in the alloy, $V_\mathrm{X}$,
which is just the change in alloy volume upon adding an atom of solute
X to the alloy, or using the concentration (or atomic fraction) of
solute X, $c_\mathrm{X}$, to yield the following: 
\be
\sizefactor = \frac{\Delta V}{V_\mathrm{ave.}} = \frac{V_\mathrm{X}-V_\mathrm{ave.}}{V_\mathrm{ave.}} = \frac{1}{V_\mathrm{ave.}}\frac{\partial V_\mathrm{ave.}}{\partial c_\mathrm{X}}.  
\ee

Our TM calculations use fixed supercells so we have determined
$\sizefactor$ by measuring the pressure, $P$, induced after
introducing a single substitutional solute into the pure solvent
metal. Any systematic and non-convergence errors in the pressure for
these large-cell calculations, which show up as a residual pressure in
the pure solvent cell calculation were subtracted in the calculation
of $P$. The volume change, $\Delta V$, associated with the introduced
solute was calculated by extrapolating to zero pressure using the bulk
modulus, $B=-V\mathrm{d}P/\mathrm{d}V$, to give
\be 
   \Delta V = \frac{PV}{B}\ \Rightarrow\ \sizefactor = \frac{NP}{B} 
\ee
where $V$ is the cell volume and $N$ is the number of atoms in the
cell, which is 256 in this case. The volume extrapolation has an
associated energy change, $E^\mathrm{corr.} = -P^2V/2B$, which, as a
result of periodic boundary condition effects, is equal to an
Eshelby-type elastic correction for a defect-containing cell embedded
in a continuous elastic medium\cite{AcklandA,HanA}. We used this
generally applicable result as a measure of the finite-volume error in
our calculations and found them to be, generally, negligible compared
to other sources of uncertainty.

\section{Results and Discussion}

\subsection{TM solutes in the defect-free lattice}
\label{defectFreeSection}

\begin{figure}[htbp]
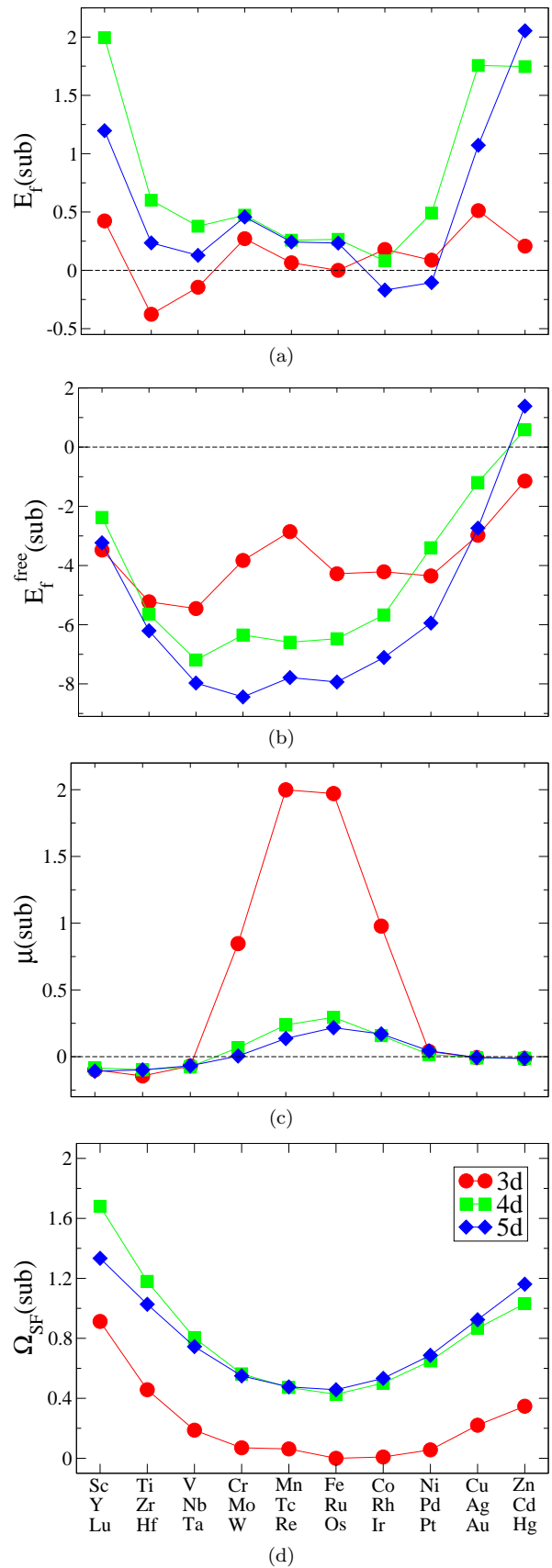

\subfigure[]{\label{EfsubAcrossGroupFig}\includegraphics[trim = 0mm 0mm 0mm 0.5mm,clip,width=0.9\columnwidth]{Fig2a_Ef-sub_across_group_vs_nd_no_x_ticks.eps}}
\subfigure[]{\label{EffreesubAcrossGroupFig}\includegraphics[trim = 0mm 0mm 0mm 0.5mm,clip,width=0.9\columnwidth]{Fig2b_Ef-free-sub_across_group_vs_nd_no_x_ticks.eps}}
\subfigure[]{\label{musubAcrossGroupFig}\includegraphics[trim = 0mm 0mm 0mm 0.5mm,clip,width=0.9\columnwidth]{Fig2c_mu-sub_across_group_vs_nd_no_x_ticks.eps}}
\subfigure[]{\label{SFAcrossGroupFig}\includegraphics[trim = 0mm 0mm 0mm 0.5mm,clip,width=0.9\columnwidth]{Fig2d_SF-sub_across_group_vs_nd.eps}}
\caption{\label{SubSolFig}Properties of a single substitutional (sub)
  solute in fct afmD Fe across the TM series: (a) sub formation
  energy, $\eformsub$, in eV ; (b) sub formation energy relative to
  the free solute atom, $\eformfreesub$, in eV ; (c) local magnetic
  moment on a sub solute atom when located in a magnetic plane with
  positive Fe moments, $\mu(\mathrm{sub})$, in $\magneton$ and (d)
  solute size factors calculated using the substitutional
  configuration, $\sfsub$. The data is given in Appendix \ref{AppB}.}
\end{figure}

We start our investigation of TM solutes in austenite with a study of
single substitutional solute properties. We present data for the
substitutional (formation) energy, magnetic moment on the
substitutional solute and solute size factor across the TM series in
\reffig{SubSolFig}. We calculate the substitutional energy relative to
the free atom, $\eformfreesub$, as well as from the standard reference
states, $\eformsub$, as in \eqn{EformEquation}, in order clarify the
discussion by removing the bias in the data coming from the varying
ground-state crystal structure. Due to the limitations of DFT
calculations, we calculate $\eformfreesub$ by subtracting the
experimental cohesive energy for the ground state crystal structure of
the solute element (at 0 K)\cite{Kittel} [see Appendix \ref{AppA}]
from $\eformsub$. Intuitively, $\eformfreesub$, is a generalisation of
the (negative) cohesive energy for the pure metals, describing the
strength of cohesion of the substitutional solute in the solvent
matrix.

The substitutional energy curve [\reffig{EfsubAcrossGroupFig}] clearly
differentiates the majority of the elements, for which $\eformsub$
lies below 0.5 eV from those elements at the extremes of the 4d and 5d
TM series (groups III, XI and XII), which exhibit substitutional
energies up to 2 eV. While no general correlation was observed with
the solute size factor, the largest solutes were also the most
insoluble. The results also show that Ti, V, Ir and Pt are readily
soluble in fct afmD Fe, which is also the case in bcc fm
Fe\cite{Olsson10}.

Changing to a free atom reference state reveals a clear parabolic
trend in $\eformfreesub$ across the series for the 4d and 5d solutes
[\reffig{EffreesubAcrossGroupFig}]. Such a trend, primarily, results
from the filling of the local d band on the solute atom as we proceed
across the series, in a similar manner to the Friedel model and its
extensions for d band cohesion in the pure transition
metals\cite{Sutton1993}. Purely atomic processes, such as the energy
needed to promote the solute atom from its electronic ground state to
that found in the metal and the loss of the atomic magnetic moment and
the associated exchange energy\cite{Brooks1983} do, however, act to
reduce this cohesion. This effect is greatest for those atoms having a
half-filled d shell and the largest atomic moments, leading to the
observed flattening of the curve near the centre of the series. While
the 3d solute data also exhibits a parabolic trend early and late in
the series, competition between these atomic processes and a lower d
band cohesion than found for the 4d and 5d solutes leads to a
pronounced reduction in solute cohesion near the centre of the
series. The competition is sufficiently strong that the elements
showing the greatest deviation from the parabolic trend, namely Cr,
Mn, Fe and Co, maintain part of their atomic moment, as can be seen in
\reffig{musubAcrossGroupFig}.

For all the other TMs, which aside from Ni have non-magnetic ground
state crystal structures, the local magnetic order in fct afmD Fe
induces small moments on the solutes. The trend in moments is similar
to that observed in bcc fm Fe\cite{Olsson10}, despite the differences
in local magnetic ordering, although the moments are much larger
there. The case of Cr is particularly interesting as it is well known
to be antiferromagnetically aligned in bcc fm Fe\cite{Olsson10} but
shows positive alignment to its magnetic plane in fct afmD Fe. We
note, however, that the nearest Fe atoms to a Cr solute in fct afmD Fe
actually lie in the adjacent and anti-aligned magnetic plane to the
one the solute is embedded in and not in the plane itself, as is the
case with all the other TM solutes. We postulate that the earlier
shift from anti-alignment to alignment, and the much smaller magnitude
moments observed in fct afmD Fe compared to bcc fm Fe, result directly
from the competing influence of these oppositely aligned 1nn Fe atoms
on the solute moment.

The size factor data [in \reffig{SFAcrossGroupFig}] exhibits a clear,
functional dependence on local d band occupancy, much as was found in
bcc Fe\cite{Olsson10}. The solute size is greatest for early and late
elements in the TM series and generally increases down a group,
although the lanthanide contraction (resulting from the weak screening
provided by the 4f shell) results in 4d and 5d solutes having similar
sizes. Size factors for a number of TM solutes have been measured
experimentally in 316L austenitic stainless
steel\cite{Straalsund1974,Kato1991}, which has an approximate
composition of Fe-17Cr-13Ni (in wt\%). We have extrapolated these
results to the case of pure Fe by assuming a fixed value for the
(partial) atomic volume of Fe atoms and compare to our work in
\reftab{sizeFactorTab}. For comparison, we also include results for
the interstitial solutes C and N from our previous
work\cite{Hepburn13}.

\begin{table}[htbp]
\begin{ruledtabular}
\begin{tabular}{ccccc}
Data & 316L                       & 316L                 & 316L steel   & This \\
Set  & steel\cite{Straalsund1974} & steel\cite{Kato1991} & extrapolated & work \\
\hline 
$V_\mathrm{ave.}$   &  11.64 &  11.60 &  11.43 & 11.14 \\
$\sizefactor$(Ti) &   ---  &  0.373 &  0.393 & 0.457 \\
$\sizefactor$(V)  &   ---  &  0.100 &  0.116 & 0.188 \\
$\sizefactor$(Cr) &  0.048 &   ---  &  0.068 & 0.070 \\
$\sizefactor$(Mn) &  0.034 &   ---  &  0.054 & 0.063 \\
$\sizefactor$(Co) & -0.065 &   ---  & -0.047 & 0.009 \\
$\sizefactor$(Ni) & -0.032 &   ---  & -0.014 & 0.056 \\
$\sizefactor$(Cu) &  0.093 &   ---  &  0.114 & 0.221 \\
$\sizefactor$(Zr) &   ---  &  1.562 &  1.600 & 1.180 \\
$\sizefactor$(Nb) &   ---  &  0.625 &  0.649 & 0.803 \\
$\sizefactor$(Mo) &  0.359 &   ---  &  0.384 & 0.563 \\
$\sizefactor$(Hf) &   ---  &  1.931 &  1.975 & 1.027 \\
$\sizefactor$(Ta) &   ---  &  0.786 &  0.813 & 0.745 \\
$\sizefactor$(C)  &  0.539 &   ---  &  0.549 & 0.529 \\
$\sizefactor$(N)  &  0.451 &   ---  &  0.460 & 0.537 \\
\end{tabular}
\end{ruledtabular}
\caption{\label{sizeFactorTab} Comparison between the average atomic
  volume per lattice site, $V_\mathrm{ave.}$, in $\angs^3$ and size
  factor, $\sizefactor$, data from this work and from experimental
  studies of 316L austenitic stainless
  steel\cite{Kato1991,Straalsund1974}. The results for C and N from
  our previous work\cite{Hepburn13} are also given. The experimental
  results have been extrapolated to the case of pure Fe by assuming a
  fixed value for the (partial) atomic volume of Fe atoms.}
\end{table}

Our calculation of the atomic volume in austenite is in good agreement
with the extrapolated experimental value, although as in bcc
Fe\cite{Olsson10} the DFT method used underestimates it by around
3\%. There is also a reasonable agreement between the size factor data
but with a general tendency of our results to overestimate the
experimental values. The underestimation of $V_\mathrm{ave.}$ is
certainly a contributing factor, although the finite experimental
temperature and the error associated with the extrapolation to pure Fe
will also contribute. For the largest solutes (Zr and Hf), however,
our results significantly underestimate the size factors. Kato {\it et
  al.}\cite{Kato1991} do, however, admit that the size factor of Hf
may well be overestimated, and the uncertainties are greatest in their
data for Hf and Zr. This may also help explain the different order of
4d and 5d solute sizes we find compared to
experiment\cite{Kato1991}. While we do agree that the group IV TMs are
larger than those in group V we find that the 4d solutes are larger
than the 5d (that is Zr$>$Hf$>$Nb$>$Ta), in contrast to Kato {\it et
  al.}  (where Hf$>$Zr$>$Ta$>$Nb) but consistent with the relative
order of atomic volumes in the pure ground state crystal structures
and with results in bcc Fe\cite{Olsson10}. Despite these
discrepancies, the generally good agreement between our results and
experiment, particularly in the reproduction of the general trend
across the TM series, gives us further confidence in our theoretical
approach to modelling austenite\cite{Klaver12,Hepburn13}.

\begin{figure}[htbp]
\subfigure[]{\label{EffreefccvsEffreebccFig}\includegraphics[trim = 0mm 0mm 0mm 0.5mm,clip,width=\columnwidth]{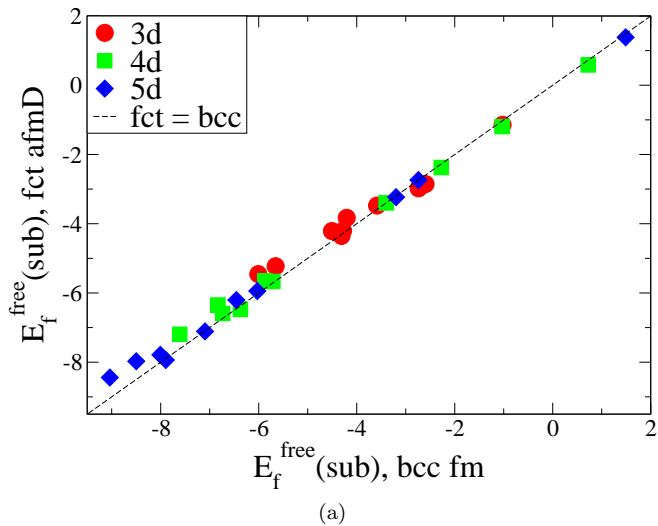}}
\subfigure[]{\label{SFfccvsSFbccFig}\includegraphics[trim = 0mm 0mm 0mm 0.5mm,clip,width=\columnwidth]{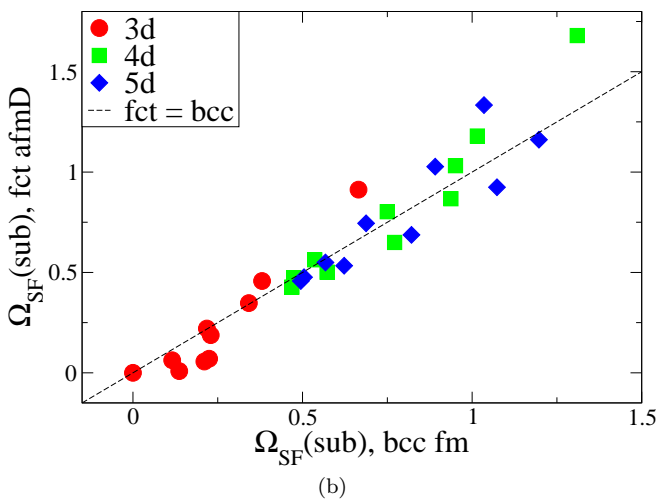}}
\caption{\label{fccvsbccsubFig}Comparison of (a) the substitutional
  formation energies relative to the free atom, $\eformfreesub$, in eV
  and (b) the solute size factors, $\sfsub$, between fct afmD Fe and
  bcc fm Fe\cite{Olsson10}.  The data is given in Appendix
  \ref{AppB}.}
\end{figure}

We have already observed a number of similarities between our results
in fct afmD Fe and those in bcc fm Fe\cite{Olsson10}. Following the
finding of strong correlation between results in pure Fe between these
two states by Klaver {\it et al.}\cite{Klaver12}, we further compare
the properties of substitutional TM solutes in the two
states. \reffig{fccvsbccsubFig} demonstrates the high level of
correlation present in the $\eformfreesub$ and $\sfsub$ data between
these two states of Fe. That said, there is a slight tendency for
solutes in the fct afmD state to exhibit greater cohesion. Overall,
these results add to the set of measurable defect and solute
properties in Fe that show a marked insensitivity to the details of
the surrounding crystal structure.

\subsection{TM Solute interactions with vacancy defects}
\label{vacancySection}

We now turn to investigate the interactions of TM solutes with
vacancies in fct afmD Fe. 

\subsubsection{Vacancy-solute binding}

\begin{figure}[htbp]
\subfigure[]{\label{EbVacSol1nnAcrossFig}\includegraphics[trim = 0mm 0mm 0mm 0.5mm,clip,width=\columnwidth]{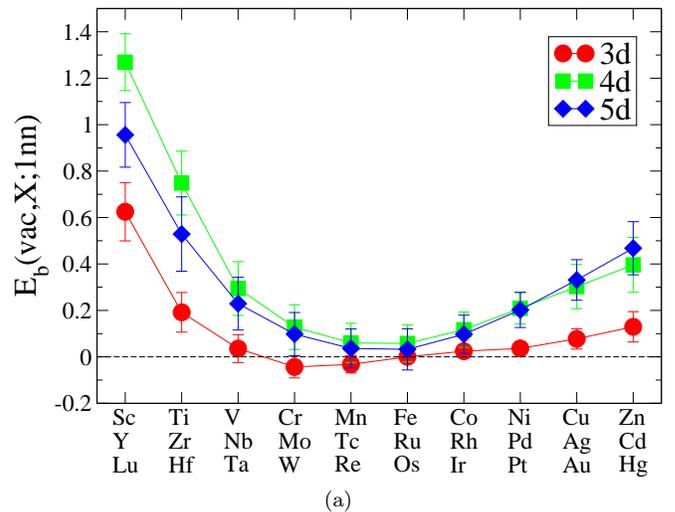}}
\vspace{12pt}
\subfigure[]{\label{EbVacSol1nnvsSFFig}\includegraphics[trim = 0mm 0mm 0mm 0.5mm,clip,width=\columnwidth]{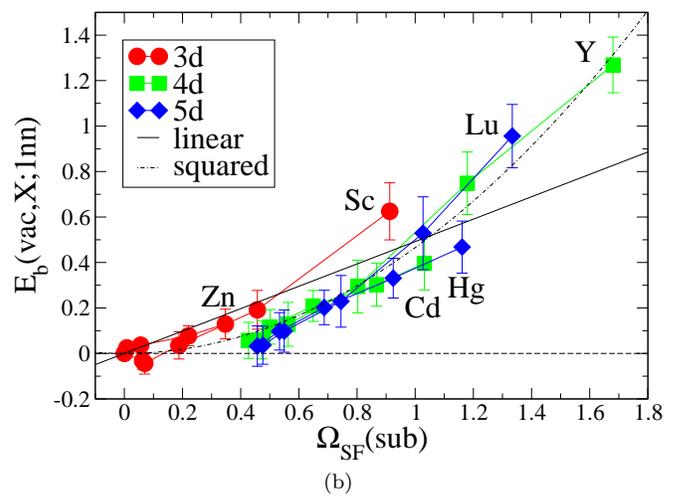}}
\caption{\label{EbVacSol1nnFig}Vacancy-solute binding energies at 1nn,
  $\ebind(\mathrm{vac,X;1nn})$, in eV (a) across the TM series and (b)
  versus the solute size factor, $\sfsub$, in fct afmD Fe. The error
  bars identify the spread in binding energies over the three distinct
  1nn sites, namely 1a, 1b and 1c in \reffig{DefectSubConfigFig}, with
  the data point chosen at the centre of this range. Figure (b) also
  shows the results of fits to the combined dataset using a linear,
  $\ebind = 0.49\ \sizefactor$, or squared, $\ebind =
  0.47\ {\sizefactor}^2$, functional dependence. The data is given in
  Appendix \ref{AppB}.}
\end{figure}

\begin{figure}[htbp]
\includegraphics[width=0.9\columnwidth]{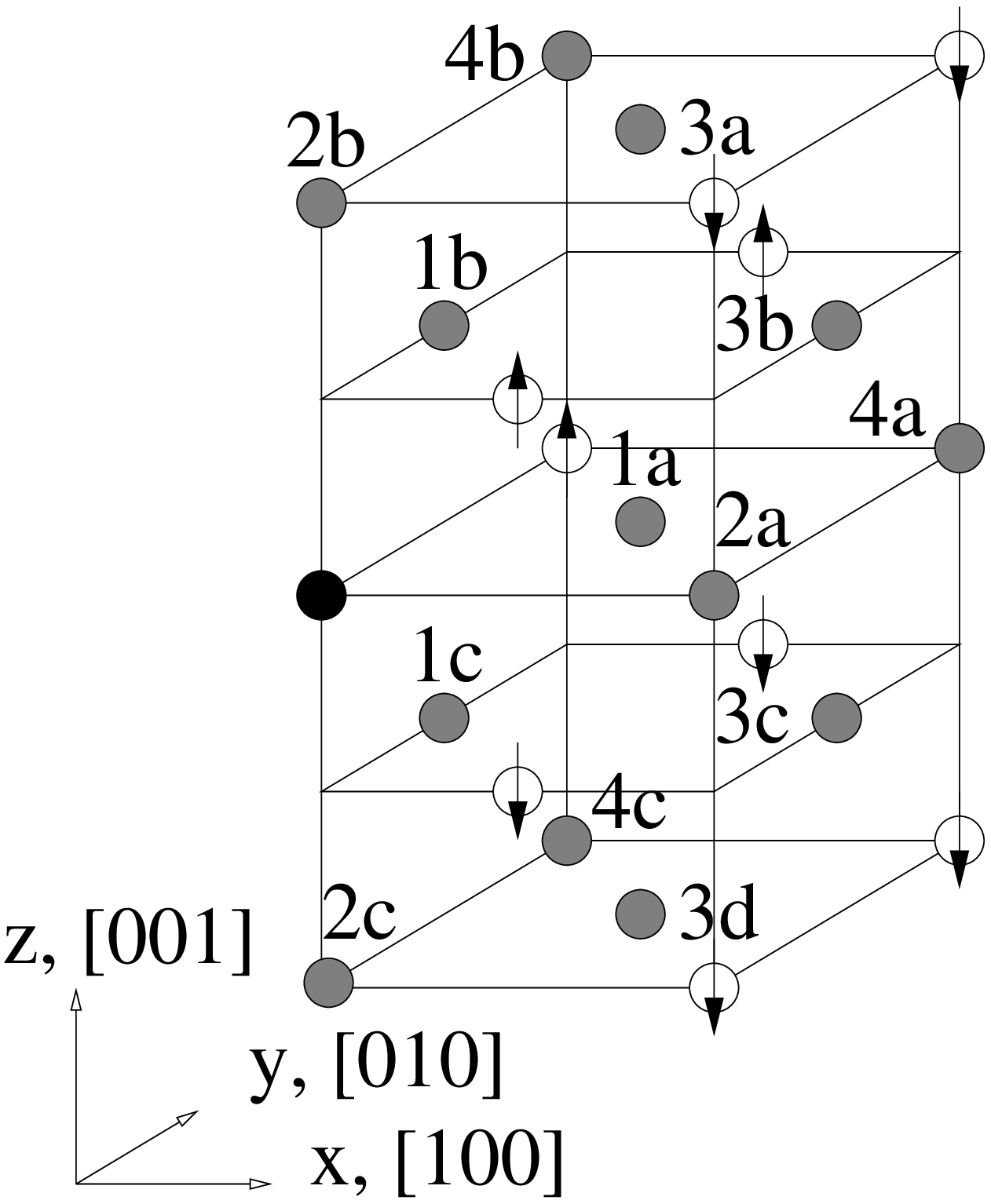}
\caption{\label{DefectSubConfigFig} Distinct configurations for an
  interacting point-defect (black) and substitutional solute (grey) in
  fct afmD Fe at up to 4nn separation. Fe atoms (white) are shown with
  arrows to indicate the local moments within their magnetic planes,
  which are also shown.}
\end{figure}

The binding energies between a vacancy and TM solute, X, at 1nn
separation, $\ebind(\mathrm{vac,X;1nn})$, are shown in
\reffig{EbVacSol1nnFig}. In fct afmD Fe, there are three distinct 1nn
configurations, labelled 1a, 1b and 1c in \reffig{DefectSubConfigFig},
and the error bars in the plots mark the spread in binding energies
with the data points chosen at the centre of the range.

The data follows a clear trend across the TM series with all elements,
aside from Cr and Mn, being attracted to the vacancy and those early
and late in the series showing the strongest binding. Experimental
estimates\cite{Kato1991} of the binding energies for Ti (0.14 eV) and
Nb (0.18 eV) in 316L steel are consistent with our data. The
similarity of the trend in the binding energy data to that for the
solute size factors [in \reffig{SFAcrossGroupFig}] is borne out in
\reffig{EbVacSol1nnvsSFFig}, which demonstrates a strong correlation
between these two quantities, although with a slight tendency for
early TMs to interact more strongly than those late in the series, as
observed in bcc Fe\cite{Olsson10}. A linear fit to the data, with a
proportionality coefficient of 0.49 eV, is close to the value of 0.45
eV found in bcc Fe\cite{Olsson10}. A function proportional to the
square of the size factor, which could be motivated from elasticity
arguments, does, however, give better agreement with the
data. Overall, these results confirm the suggestions from
experiment\cite{Kato1991,Kato1992} and theory\cite{Stepanov2004} that
oversized solutes act as trapping sites for vacancies.

What is not apparent from \reffig{EbVacSol1nnFig} is that the largest
solutes, namely Sc, Y, Zr, Lu and Hf, relax to exactly half way
between their original lattice site and the vacancy at 1nn, that is to
the centre of the associated divacancy, forming what we refer to as a
solute-centred divacancy (SCD). All other solutes remain on-site
during relaxation. This behaviour was already observed for He in
austenite\cite{Hepburn13} and for the same TM solutes in bcc
Fe\cite{Olsson10} and clearly has important implications for
vacancy-mediated solute diffusion, which we now discuss.

\subsubsection{Vacancy-mediated solute diffusion}

\begin{figure}[htbp]
\includegraphics[width=0.8\columnwidth]{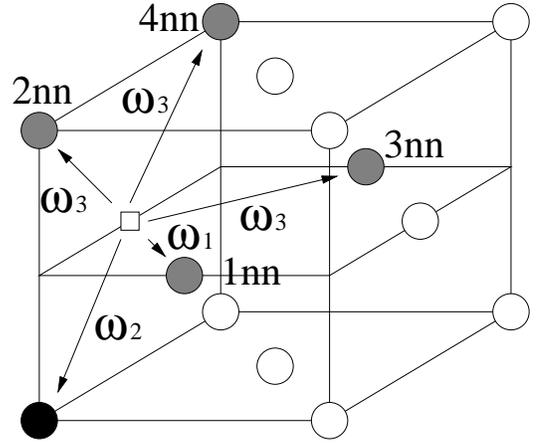}
\caption{\label{fiveFreqFig} The distinct types of vacancy (white
  square) jumps near a substitutional solute (black circle) in the fcc
  lattice for the five-frequency model of Lidiard and
  LeClaire\cite{Lidiard55,LeClaire56}. Solvent metal atoms involved in
  the vacancy jumps (grey circles) are distinguished from those in the
  background matrix (white circles). With the vacancy initially at 1nn
  to the solute the jumps can either maintain a 1nn separation
  ($\omega_1$), have the vacancy exchange with the solute ($\omega_2$)
  or involve dissociation to ($\omega_3$) or association from
  ($\omega_4$) 2nn, 3nn and 4nn separation. In the model, all other
  vacancy jumps are considered identical to that in the pure solvent
  metal ($\omega_0$).}
\end{figure}

The vacancy-mediated diffusion of a substitutional solute in an fcc
lattice is usually well described by the five-frequency model of
Lidiard and LeClaire\cite{Lidiard55,LeClaire56}. The distinct types of
vacancy jumps, as labelled by their associated frequencies,
$\omega_i$, are given in \reffig{fiveFreqFig}.

\begin{figure}[htbp]
\includegraphics[trim = 0mm 0mm 0mm 0.5mm,clip,width=\columnwidth]{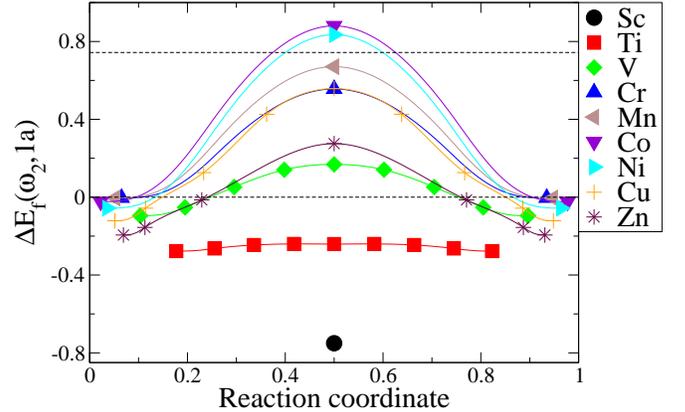}
\caption{\label{vacSolExchFig}Change in formation energy,
  $\Delta\eform(\omega_2,\mathrm{1a})$, in eV, for the 3d TM solutes
  in fct afmD Fe along the 1a jump path for vacancy-solute exchange
  [see \reffig{DefectSubConfigFig}]. The zero of energy corresponds to
  a non-interacting vacancy and substitutional solute. The reaction
  coordinate is the solute position, after rescaling, with 0 or 1
  corresponding to a perfectly on-site solute and 0.5 to the case
  where it is half-way between the two lattice sites, that is to the
  SCD. The higher dotted line gives the vacancy migration energy for
  this jump path in pure fct afmD Fe.}
\end{figure}

The frequencies are related to migration barriers by Arrhenius-type
expressions,
\be\label{omegaEquation}
   \omega_i = C_{\mathrm{m},i}\exp(-\beta \emig(\omega_i)), 
\ee 

where $\beta = 1/k_\mathrm{B}T$ and $\emig(\omega_i)$ is the migration
energy for the jump. For vacancies in an fcc lattice, the single
maximum in energy along the jump path [see \reffig{vacSolExchFig}]
defines the transition state (TS) and $\emig(\omega_i)$ is, therefore,
the energy difference between the TS and the initial jump
configuration. A nearby solute, X, can change the energy of both of
these configurations (relative to a non-interacting state). For the
initial configuration, I, this is quantified by the vacancy binding
energy, $\ebind(\mathrm{vac,X;I})$, and we can, similarly, define a
``binding energy to the transition state'',
$\ebind(\mathrm{TS,X};\omega_i)$. The change in migration energy
relative to that in pure Fe is then given by

\be 
\label{emigEquation}
   \emig(\omega_i) = \emig(\omega_0) + \ebind(\mathrm{vac,X;I}) - \ebind(\mathrm{TS,X};\omega_i).
\ee

We first investigate vacancy-solute exchange, that is jump $\omega_2$,
for which there are three distinct paths in fct afmD Fe [see
  \reffig{DefectSubConfigFig}], namely 1a, 1b and
1c. \reffig{vacSolExchFig} shows the change in formation energy along
the 1a jump path, $\Delta\eform(\omega_2,\mathrm{1a})$, for the 3d
solutes. All solutes relax towards a vacancy at 1nn, with larger
solutes relaxing further, and Sc going to the symmetric position to
form a stable SCD configuration, which is the TS for the other
solutes. While the increasing vacancy binding energy leads to a steady
lowering of the initial on-site energy, the TS binding energy
increases more quickly with size factor, leading to a net lowering of
the migration barrier, which is ultimately responsible for the
formation of the stable SCD for Sc. Similar results were found for the
other migration paths and TM solutes, with Sc, Y, Lu, Zr and Hf
forming stable SCDs. 

\begin{figure}[htbp]
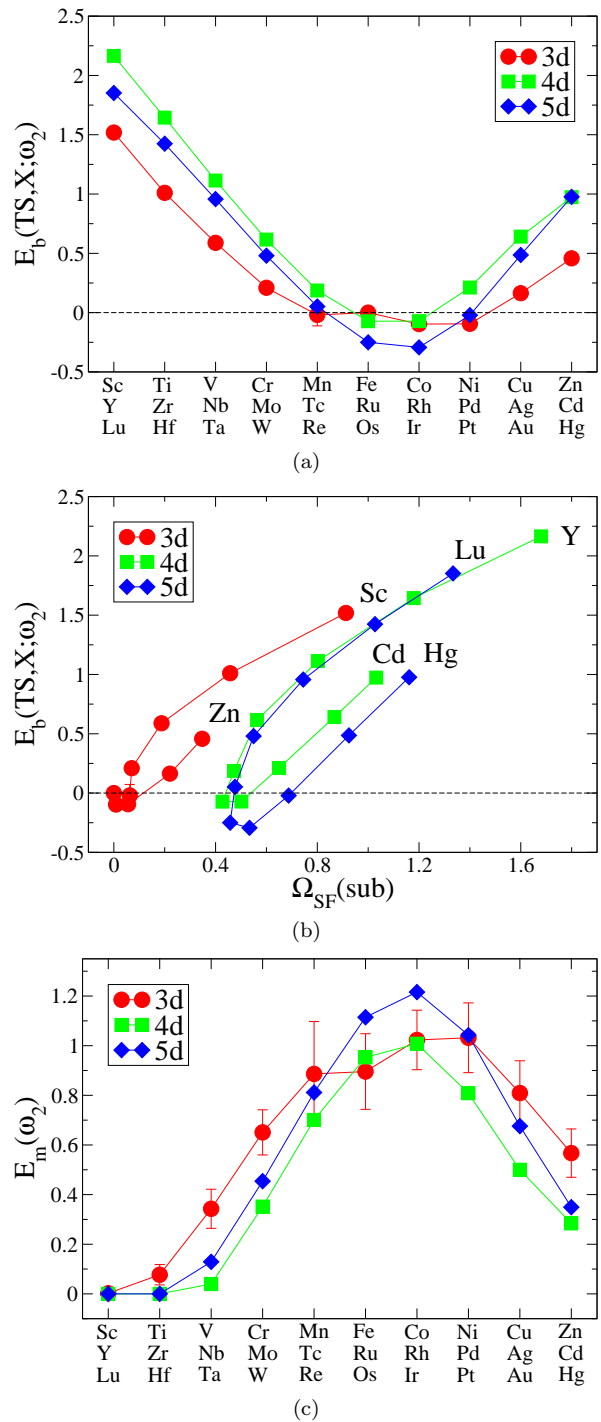

\subfigure[]{\label{EbVacSolExchTS1nnAcrossFig}\includegraphics[trim = 0mm 0mm 0mm 0.5mm,clip,width=0.9\columnwidth]{Fig8a_Eb-vac-sol-exch-TS-1nn-ave_across_group_vs_nd.eps}}
\subfigure[]{\label{EbVacSolExchTS1nnvsSFFig}\includegraphics[trim = 0mm 0mm 0mm 0.5mm,clip,width=0.9\columnwidth]{Fig8b_Eb-vac-sol-exch-TS-1nn-ave_vs_SF-sub.eps}}
\subfigure[]{\label{EmVacSolExch1nnAcrossFig}\includegraphics[trim = 0mm 0mm 0mm 0.5mm,clip,width=0.9\columnwidth]{Fig8c_Em-vac-sol-1nn-exch-ave_across_group_vs_nd.eps}}
\caption{\label{EmEbVacSolExch1nnFig}Migration energies,
  $\emig(\omega_2)$, and binding energies to the transition state,
  $\ebind(\mathrm{TS,X};\omega_2)$, in eV for vacancy-solute exchange
  in fct afmD Fe. The error bars for the 3d solutes and for Y, Zr, Lu
  and Hf show the spread in energies over the 1a and 1b jump paths
  [see \reffig{DefectSubConfigFig}] with the data point taken as the
  mean value. The data for path 1c was deemed unreliable given that
  the magnetic moment on the migrating atom is constrained to be zero
  in the transition state, leading to an overestimation of the
  migration energy\cite{Klaver12}. Systematic cancellations have
  resulted in the error bars being smaller than the symbols in figures
  (a) and (b). For the other solutes only the 1a jump data is
  shown. The data is given in Appendix \ref{AppB}.}
\end{figure}

\reffig{EmEbVacSolExch1nnFig} shows $\emig(\omega_2)$ and
$\ebind(\mathrm{TS,X};\omega_2)$ across the TM series. The TS binding
energy trend shows strong positive binding at the beginning and end of
the series. While there is no simple relationship between solute size
and TS binding [\reffig{EbVacSolExchTS1nnvsSFFig}], in contrast to the
vacancy binding [\reffig{EbVacSol1nnvsSFFig}], the two are still
strongly correlated. Generally, the TS binding energy grows at a
greater rate than the vacancy binding energy with size factor, which
\eqn{emigEquation} shows leads to an overall reduction in
$\emig(\omega_2)$ as we move out from the centre of the TS series
    [\reffig{EmVacSolExch1nnAcrossFig}]. The extreme examples are Sc,
    Y, Zr, Lu and Hf, where the energy barrier ceases to exist and the
    SCD is stable. The barrier heights for Ti, Nb and Ta are also
    effectively negligible [see \reffig{vacSolExchFig} for Ti] and
    should be considered as forming stable SCDs at finite
    temperature. Near the centre of the series, by contrast, a
    combination of positive binding to the vacancy and negative
    binding to the transition state (see Os and Ir in particular)
    leads to greater migration energies than in pure Fe. It is also
    interesting to note that the significant difference between the
    $\omega_2$ jumps for Cr and Ni found previously\cite{Klaver12},
    predominantly result from differences in binding to the transition
    state (instead of the vacancy), which results, most likely, from
    magnetic interactions, given the similar solute sizes.

We also investigated the relative importance of vacancy-solute
exchange at 2nn to vacancy-mediated diffusion using Y in fct afmD
Fe. This was motivated by results from our previous work on
substitutional He\cite{Hepburn13}. While the TS binding energy for Y
along the 2a jump path [see \reffig{DefectSubConfigFig}] was
significant at 1.82 eV, it was only sufficient to reduce the migration
energy to 1.74 eV. Using the data for $\omega_2$ jumps as a reference
[see \reffig{EmEbVacSolExch1nnFig}] we would expect the migration
barriers for the other TM solutes to be in excess of the Y value and
can, therefore, conclude that vacancy-solute exchange at 2nn is
unlikely to contribute significantly to their vacancy-mediated
diffusion.

\begin{figure}[htbp]
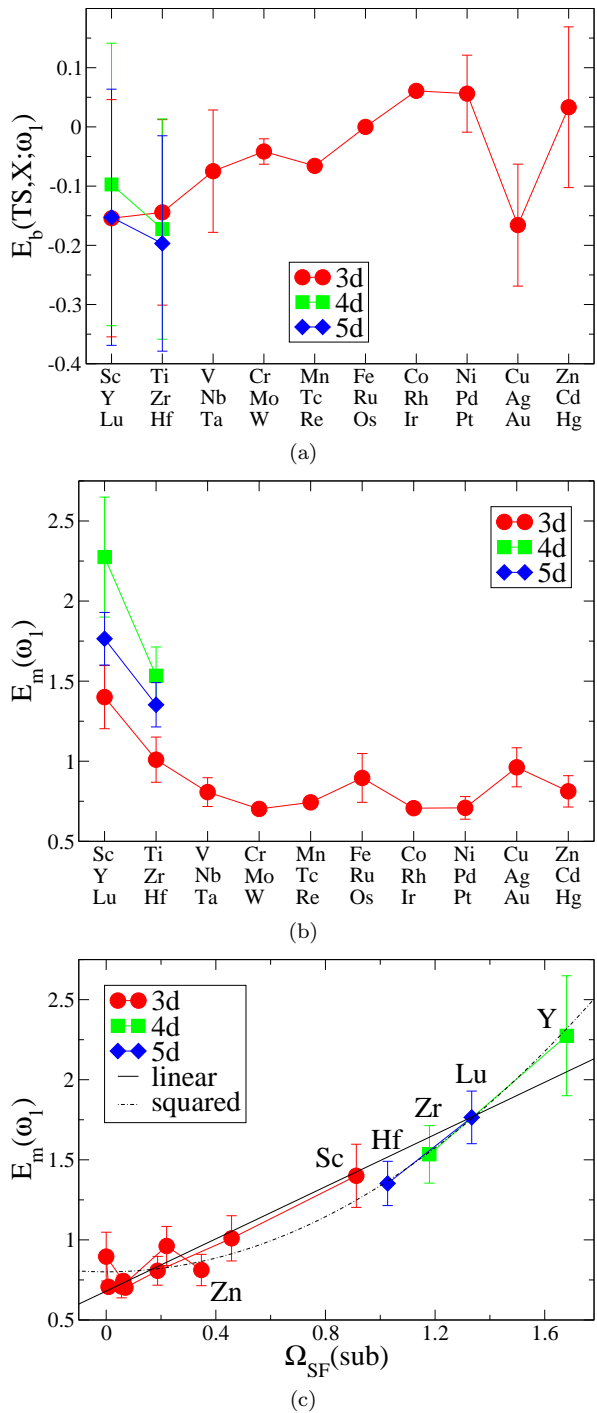

\subfigure[]{\label{EbTSSolw1AcrossFig}\includegraphics[trim = 0mm 0mm 0mm 0.5mm,clip,width=0.9\columnwidth]{Fig9a_Eb-vac-sol-w1-TS-ave_across_group_vs_nd.eps}}
\subfigure[]{\label{Emw1AcrossFig}\includegraphics[trim = 0mm 0mm 0mm 0.5mm,clip,width=0.9\columnwidth]{Fig9b_Em-vac-sol-w1-ave_across_group_vs_nd.eps}}
\subfigure[]{\label{Emw1vsSFFig}\includegraphics[trim = 0mm 0mm 0mm 0.5mm,clip,width=0.9\columnwidth]{Fig9c_Em-vac-sol-w1-ave_vs_SF-sub.eps}}
\caption{\label{w1Fig}Migration energies, $\emig(\omega_1)$, and
  binding energies to the transition state,
  $\ebind(\mathrm{TS,X};\omega_1)$, in eV for $\omega_1$ jumps in fct
  afmD Fe. For all solutes except Y we have considered the jumps from
  1b to 1b and from 1c to 1c configurations [see
    \reffigs{DefectSubConfigFig}{fiveFreqFig}], which have
  symmetry-stabilised transition states, and the error bars reflect
  the spread in these values. For the special case of Y the 1a to 1b
  jump, which required the use of climbing image NEB calculations, was
  also included. The 1a to 1c jump was excluded from the analysis
  given that the constraints of collinear spin calculations would lead
  to a zero moment on the migrating Fe atom at some point on the path,
  resulting in a significant overestimation of the migration
  energy. Figure (c) also shows the results of linear, $\emig = 0.68 +
  0.81\ \sizefactor$, or squared, $\emig = 0.80 +
  0.54\ {\sizefactor}^2$, fits to the combined data. The data is given
  in Appendix \ref{AppB}.}
\end{figure}

For the $\omega_1$ jumps, we have focussed on the 3d solutes and those
that form stable SCDs, namely Y, Zr, Lu and Hf. The results are
presented in \reffig{w1Fig}. While the dependence on details of local
magnetic state for some elements is large, the TS binding energy is,
generally, negative and much smaller than either the $\omega_2$ TS or
1nn vacancy binding energies. Both the trend in the $\emig(\omega_1)$
data [\reffig{Emw1AcrossFig}] and its correlation with the size factor
[\reffig{Emw1vsSFFig}], therefore, primarily result from the
vacancy-solute binding energy data [\reffig{EbVacSol1nnFig}]. The
intuitive result is that the migration energy for an $\omega_1$ jump
increases with solute size factor and the data in \reffig{Emw1vsSFFig}
is well described by a linear or quadratic fit function.

\begin{figure}[htbp]
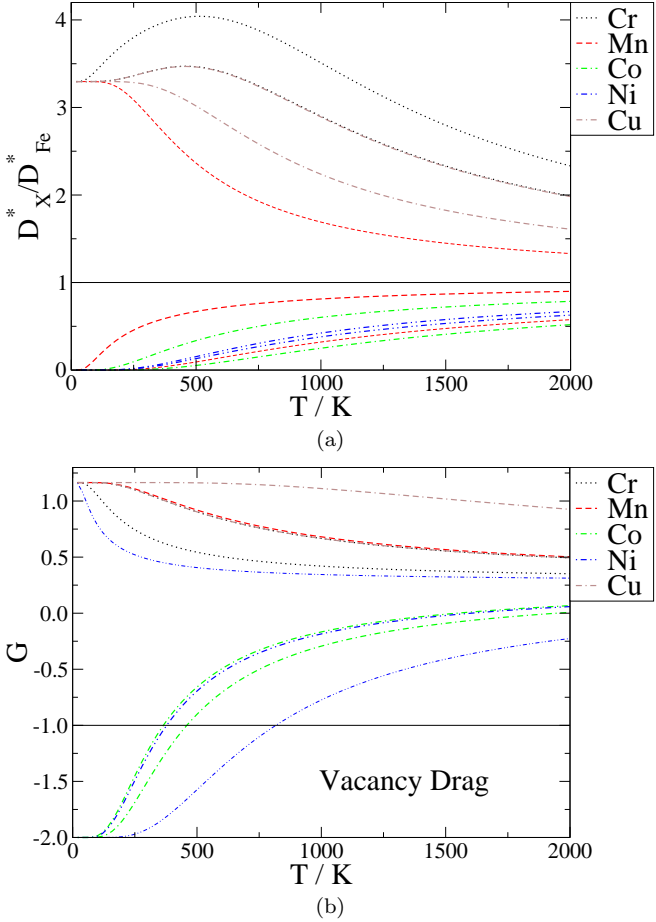

\subfigure[]{\label{RatioFig}\includegraphics[trim = 0mm 0mm 0mm 0.5mm,clip,width=\columnwidth]{Fig10a_RXFe_vs_T_low_high.eps}}
\subfigure[]{\label{WindyFig}\includegraphics[trim = 0mm 0mm 0mm 0.5mm,clip,width=\columnwidth]{Fig10b_GX_vs_T_low_high.eps}}
\caption{\label{DiffusionFig}Temperature dependence of (a) the tracer
  diffusion coefficient ratio, $D^*_\mathrm{X}/D^*_\mathrm{Fe}$, and
  (b) the vacancy wind parameter, $G$, for the solutes Cr (black,
  dotted line), Mn (red, dashed line), Co (green, dot-dashed line), Ni
  (blue, double-dot-dashed line) and Cu(brown, dot-double-dashed
  line). The lower and upper curves for each element show the likely
  uncertainties. The third, bolder, central curves for Mn in figure
  (a) and Ni in figure (b) correspond to the central parameter values
  in \reftab{EbTSTab}. The vacancy drag regime ($G<-1$) is identified
  in figure (b).}
\end{figure}

In previous analysis of Ni and Cr\cite{Klaver12}, the ratio of the
tracer diffusion coefficient for solute X to that for the solvent,
$D^*_\mathrm{X}/D^*_\mathrm{Fe}$, and the vacancy wind parameter, $G$,
which measures the relative orientations of the solute and vacancy
fluxes, were found to be controlled by only two parameters, namely
$\ebind(\mathrm{TS,X};\omega_1)$ and
$\ebind(\mathrm{TS,X};\omega_2)$. By using the fact that
$D^*_\mathrm{X}/D^*_\mathrm{Fe}$ is a monotonically increasing
function of both parameters and $G$ is a monotonically decreasing
function of $\ebind(\mathrm{TS,X};\omega_1)$, we have plotted lower
and upper bounds for these quantities for Cr, Mn, Co, Ni and Cu in
\reffig{DiffusionFig} using the parameter values and their
uncertainties in \reftab{EbTSTab}.

\begin{table}[htbp]
\begin{ruledtabular}
\begin{tabular}{crr}
 Element & $\ebind(\mathrm{TS,X};\omega_1)$ & $\ebind(\mathrm{TS,X};\omega_2)$ \\
\hline
Cr & $-0.042\pm 0.022$ &  $0.210\pm 0.022$ \\
Mn & $-0.066\pm 0.000$ & $-0.019\pm 0.092$ \\
Co &  $0.061\pm 0.007$ & $-0.097\pm 0.040$ \\
Ni &  $0.056\pm 0.065$ & $-0.094\pm 0.003$ \\
Cu & $-0.166\pm 0.103$ &  $0.164\pm 0.022$ \\
\end{tabular}
\end{ruledtabular}
\caption{\label{EbTSTab} Binding energies to the transition state for
  $\omega_1$ and $\omega_2$ jumps, in eV. The errors and central
  values are as described in \reffigs{EmEbVacSolExch1nnFig}{w1Fig}.}
\end{table}

We find that Ni and Co diffuse at a similar rate and more slowly than
Fe, primarily as a result of their negative binding to the $\omega_2$
transition state. In contrast, both Cu and Cr exhibit positive TS
binding and diffuse more quickly than Fe. For Mn,
$\ebind(\mathrm{TS,X};\omega_2)$ spans both positive and negative
values, leading to uncertainty in
$D^*_\mathrm{X}/D^*_\mathrm{Fe}$. The data does, however, show that Mn
will diffuse more slowly than either Cr or Cu. In addition to this
analysis of diffusion rates, the vacancy wind parameter, $G$, allows
one to determine whether the solute flux induced by vacancy-mediated
diffusion would be in the same direction (for $G<-1$) or opposite to
(for $G>-1$) any vacancy flux. The net diffusion of Cr, Mn and Cu is
opposite to the vacancy flux at all temperatures
[\reffig{WindyFig}]. While the behaviour of Ni does appear poorly
determined, this results from the $\ebind(\mathrm{TS,X};\omega_1)$
parameter extending to the small but negative value of -0.009 eV at
its lower bound, leading to the divergent behaviour seen between the
upper and two lower curves for $G$. A positive value is much more
likely, meaning that the Ni flux flips from being opposite to the
vacancy flux to being in the same direction below a critical
temperature, $T_\mathrm{c}$, where radiation-induced vacancies drag
the solutes with them to the defect sinks. We estimate a value of
$T_\mathrm{c}=400\pm 50$ K for Co. Overall, these observations are
consistent with the RIS of Cr away from and Ni towards vacancy sinks
in austenitic stainless
steels\cite{Kato1991,Kato1992,Allen2005}. Furthermore, they suggest
that Co concentrations will be enhanced and Cu depleted from vacancy
sinks. The behaviour of Mn remains undetermined in this study as it
depends critically on whether it diffuses faster or slower than Fe,
leading, respectively, to depletion or enhancement at defect sinks.

Another useful area of approximation is the case where the $\omega_2$
jump frequency becomes very much greater than both $\omega_1$ and
$\omega_3$. This approximation not only applies when $\emig(\omega_2)$
is small, as is the case for many oversized solutes, but also allows
us to treat the case when the migration barrier ceases to exist and a
stable SCD is formed.  In this limit the general expression for
$D^*_\mathrm{X}$ [see Klaver {\it et al.}\cite{Klaver12}] becomes
independent of $\omega_2$ and is given by,
\be\label{DstarEquation}
D^*_\mathrm{X} = a^2 c_\mathrm{V} C_\mathrm{b}\exp(\beta \ebind(\mathrm{vac,X;1nn}))\left[ \omega_1 + \frac{7}{2}F(\frac{\omega_4}{\omega_0})\omega_3\right], 
\ee 
where $a$ is the fcc lattice parameter, $c_\mathrm{V}$, is the vacancy
concentration, $C_\mathrm{b}$ is a weakly temperature-dependent
prefactor that depends on the vacancy-solute binding entropy and the
function, F, gives the fraction of dissociative ($\omega_3$) jumps
that do not effectively return the vacancy to its original
site\cite{Manning}.

The physical interpretation of the large $\omega_2$ limit is that the
solute oscillates rapidly over a small $\omega_2$ barrier or is
located about the centre of the associated divacancy, until an
$\omega_1$ or $\omega_3$ jump takes place. $\omega_1$ corresponds to
the migration of the (effective) SCD as a single entity, which we
investigated as a primary mechanism for substitutional He diffusion
previously\cite{Hepburn13}. $\omega_3$ corresponds to the net
diffusion resulting from dissociation (and reassociation) events. The
activation energy for both of these diffusion mechanisms is given by
\bea \label{EAEquation} 
E_\mathrm{A}(\omega_i) & = & \emig(\omega_i) - \ebind(\mathrm{vac,X;1nn})\ \left[\ + \eform(\mathrm{vac})\ \right]\nonumber \\
& = & \emig(\omega_0) - \ebind(\mathrm{TS,X};\omega_i)\ \left[\ + \eform(\mathrm{vac})\ \right],
\eea
where the vacancy formation energy, $\eform(\mathrm{vac})$, is either
present for a thermal vacancy population or absent for a fixed
supersaturation of vacancies, as found in irradiated materials, and
the tracer diffusion coefficient remains proportional to the vacancy
concentration. \eqn{EAEquation} shows that $E_\mathrm{A}(\omega_i)$ is
lower than the activation energy for (tracer) self-diffusion by the TS
binding energy, $\ebind(\mathrm{TS,X};\omega_i)$. We note, in passing,
that while we did not consider the $\omega_3$ diffusion mechanism for
substitutional He previously\cite{Hepburn13}, test calculations showed
it should exhibit a similar TS binding and activation energy to the
$\omega_1$ mechanism.

For the TS solutes the $\ebind(\mathrm{TS,X};\omega_1)$ data in
\reffig{EbTSSolw1AcrossFig} suggests that the activation energy for
the $\omega_1$ diffusion mechanism will, generally, be higher than for
self-diffusion. A general study of $\omega_3$ (and $\omega_4$) jumps
would have been prohibitively expensive, given the requirement of 10
relaxed configuration calculations and 9 NEB calculations per
solute. We have, however, completed this study for Y, both as the
largest solute and for its importance in ODS steels. The results are
summarised in \reftab{EmvacYTab} along with suitably-averaged
effective values for the $\omega_3$ and $\omega_4$ jump data following
the method of Tucker {\it et al.}\cite{Tucker10}.

\begin{table}[htbp]
\begin{ruledtabular}
\begin{tabular}{crrr}
 & $\emig(\omega_i)$ & $\ebind(\mathrm{vac,X;I})$ & $\ebind(\mathrm{TS,X};\omega_i)$ \\
\hline
$\omega_0$      & $0.90\pm 0.15$ & 0 & 0 \\
$\omega_1$      & $2.27\pm 0.37$ &  $1.27\pm 0.12$ & $-0.10\pm 0.24$ \\
$\omega_2$      & 0              &  $1.27\pm 0.12$ &  $2.16\pm 0.03$ \\
$\omega_3$, 2nn & $1.97\pm 0.16$ &  $1.27\pm 0.12$ &  $0.28\pm 0.22$ \\
$\omega_3$, 3nn & $1.43\pm 0.23$ &  $1.27\pm 0.12$ &  $0.65\pm 0.20$ \\
$\omega_3$, 4nn & $1.31\pm 0.05$ &  $1.27\pm 0.12$ &  $0.93\pm 0.05$ \\
$\omega_3$, eff & $1.40\pm 0.20$ &                 &  $0.76\pm 0.23$ \\
$\omega_4$, 2nn & $0.59\pm 0.28$ & $-0.13\pm 0.03$ &  $0.28\pm 0.22$ \\
$\omega_4$, 3nn & $0.20\pm 0.12$ &  $0.09\pm 0.08$ &  $0.65\pm 0.20$ \\
$\omega_4$, 4nn & $0.13\pm 0.04$ &  $0.15\pm 0.07$ &  $0.93\pm 0.05$ \\
$\omega_4$, eff & $0.20\pm 0.12$ &                 &  $0.76\pm 0.23$ \\
\end{tabular}
\end{ruledtabular}
\caption{\label{EmvacYTab} Migration energy, $\emig(\omega_i)$, and
  solute binding energies to the vacancy in the initial jump
  configuration, $\ebind(\mathrm{vac,X;I})$, and the transition state,
  $\ebind(\mathrm{TS,X};\omega_i)$, in eV for vacancy jumps near a Y
  solute in fct afmD Fe. The distinct jumps are given in
  \reffig{fiveFreqFig}. Note that there is only one transition state
  for corresponding $\omega_3$ and $\omega_4$ jumps so the binding
  energies are identical. The errors give the spread in energies over
  the distinct $\omega_i$ jump paths in fct afmD Fe or initial
  configurations [see \reffig{DefectSubConfigFig}] with the given
  value chosen at the centre of this range. The data is given in
  Appendix \ref{AppB}. The effective (eff) $\omega_3$ and $\omega_4$
  migration barriers (and TS binding energies) were calculated from
  those for 2nn, 3nn and 4nn jumps following the method of Tucker {\it
    et al.}\cite{Tucker10} and are valid in a temperature range from 0
  to 2000 K.}
\end{table}

The vacancy binding energy data exhibits a clear trend of strong
attraction at 1nn followed by a much weaker repulsive interaction at
2nn and weak attraction at 3nn and 4nn separations. The same trend was
reported in fcc nm Fe\cite{Gopejenko}, although with a discrepancy in
binding energy of up to 0.4 eV. We put this discrepancy down to their
choice of much smaller (96 atom) supercells rather than the difference
in magnetic reference state, given that our own (256 atom cell)
calculations in fcc nm Fe found binding energies at the centres of the
ranges reported in \reftab{EmvacYTab}. It is interesting to note that
a very similar trend was also observed for early TM solutes in bcc
Fe\cite{Olsson10} and for He in austenite\cite{Hepburn13}. In contrast
to the rather sharp fall-off in vacancy binding, the binding energies
to the $\omega_3$ (and $\omega_4$) transition states remain high at as
much as 1 eV. This translates into much lower migration energies than
for $\omega_1$ jumps and the $\omega_3$ mechanism will, therefore,
dominate the vacancy-mediated diffusion of Y solutes. While the
$\omega_3$ migration energies remain greater than in pure Fe, the high
TS binding energies mean that the corresponding activation energies
[\eqn{EAEquation}] are much lower than for self-diffusion. This result
means that Y will diffuse faster than Fe above some critical
temperature, despite its much greater size. Another important
consequence of the strong TS binding energies are the very low
migration energies for $\omega_4$ jumps, with an effective value of
$0.20\pm 0.12$ eV, which is significantly less than in pure Fe. Such a
low value means that a newly dissociated vacancy is much more likely
to return to the solute than be lost to the general matrix (making an
$\omega_0$ jump) and it is reasonable to ask why this does not
significantly suppress diffusion through the factor,
$F(\omega_4/\omega_0)$ in \eqn{DstarEquation}. However, even in the
limit where the vacancy always returns, that is $\omega_4/\omega_0
\rightarrow +\infty$, $F$ remains above zero at $2/7$. This results
from the fact that the vacancy can return to different sites at 1nn to
the solute from the one it left and, therefore, still contribute to
diffusion\cite{Manning}.

Thus we can state with reasonable confidence that the $\omega_3$
diffusion mechanism will dominate for the early (oversized) TM solutes
with an activation energy lower than that for self-diffusion. The
enhanced mobility of these solutes is, certainly, an important factor
in understanding the nucleation and formation of the complex oxide
nanoparticles produced during the manufacturing of ODS steels,
although other factors, such as oxygen mobility and the interactions
between the oxide components will also be important\cite{Gopejenko}.

\subsubsection{Vacancy clustering and void nucleation}

In the experimental work of Kato {\it et al.}\cite{Kato1991} it was
shown that, while the addition of oversized TM solutes to 316L steel
did suppress void growth and reduce void swelling under irradiation,
this was accompanied by an abrupt increase in void number density
above a certain solute size factor and for sufficiently-high radiation
doses. On this basis, the authors suggested that the oversized solutes
were acting as void nucleation sites. We have investigated this
possibility by studying the growth of vacancy clusters around a single
Y atom, that is vac$_n$-Y clusters, and the binding energies for the
most stable clusters are given in \reftab{vacYClusterTab}.

\begin{table}[htbp]
\begin{ruledtabular}
\begin{tabular}{cccccc}
 $n$ & 1 & 2 & 3 & 4 & 5 \\
\hline
$\ebind^\mathrm{tot}$ & $1.27\pm 0.12$ & $3.12\pm 0.01$ & $5.11\pm 0.02$ & 5.06 & 6.51 \\
$\ebind^\mathrm{vac}$ & $1.27\pm 0.12$ & $1.85\pm 0.14$ & $1.99\pm 0.03$ & $-0.05\pm 0.02$ & 1.46 \\
$\ebind^\mathrm{Y}$   & $1.27\pm 0.12$ & $3.02\pm 0.08$ & $4.50\pm 0.05$ & $4.06\pm 0.29$ & 4.83 \\
\end{tabular}
\end{ruledtabular}
\caption{\label{vacYClusterTab} The total binding energy between $n$
  vacancies and a substitutional Y solute in a vac$_n$-Y cluster,
  $\ebind^\mathrm{tot}$, the binding energy for adding a vacancy to a
  vac$_{n-1}$-Y cluster, $\ebind^\mathrm{vac}$, and the binding energy
  of a substitutional Y solute to a vac$_n$ cluster,
  $\ebind^\mathrm{Y}$, in eV for the most stable clusters in fct afmD
  Fe. The difference between $\ebind^\mathrm{tot}$ and
  $\ebind^\mathrm{Y}$ is, therefore, the total binding energy of the
  most stable vac$_n$ cluster. The errors give the spread in binding
  energies over the distinct configurations in fct afmD Fe that would
  be equivalent in austenite with the data point taken at the centre
  of this range. For $n=4$, only the most stable configuration was
  used and for $n=5$ the most stable cluster is uniquely defined in
  fct afmD Fe.}
\end{table}

A single vacancy binds strongly at 1nn to a Y solute, forming a stable
SCD, as discussed previously. Once formed, an SCD acts as an even
stronger trap for vacancies than the Y solute alone. A second vacancy
binds to the SCD to form a close-packed triangle of vacancies lying in
a (111) plane with the Y atom at the centre. The corresponding
configuration in pure fct afmD Fe was found to be unstable, suggesting
that the configuration is only stable for solutes above a critical
size, as observed for the SCD. The pattern continues with the addition
of a third vacancy, which binds to form a tetrahedron of vacancies,
mutually at 1nn separation, with the Y atom, once again, relaxing to
the centre of this proto-void. This type of configuration, which in
pure fcc metals is know as the Damask-Dienes-Weizer (DDW)
structure\cite{Damask1959} and is the smallest possible stacking fault
tetrahedron\cite{Vineyard1961} there, was also found to be the most
stable trivacancy cluster in austenite\cite{Klaver12}.

\begin{figure}[htbp]
\includegraphics[width=0.8\columnwidth]{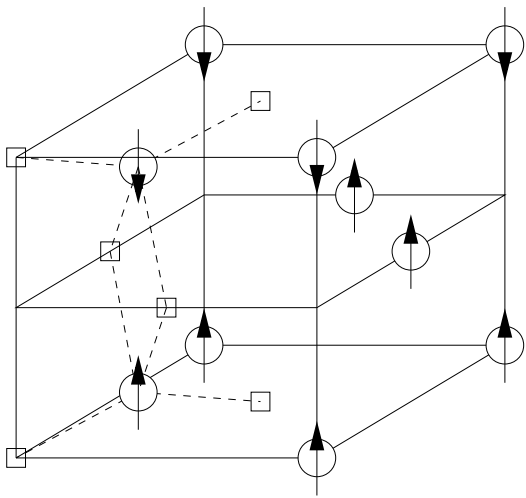}
\caption{\label{stackedDDWFig}The most stable of the three distinct
  stacked-DDW structures for a tetravacancy cluster in fct afmD
  Fe. The arrows indicate the local moments on the Fe atoms (circles)
  and the magnetic planes are shown explicitly. Vacancies are shown as
  small squares. The two central Fe atoms repel one another away from
  their relaxed positions in the DDW sub-units but maintain the large
  moments of around $3\ \magneton$ found previously\cite{Klaver12}. }
\end{figure}

Following work in fcc Al\cite{Wang2011} we considered the natural
growth of these highly-stable DDW-type structures by investigating
tetravacancy clusters of the form shown in \reffig{stackedDDWFig},
which we refer to as stacked-DDW structures. In pure Fe we found them
to be more stable than all other tetravacancy clusters considered
previously\cite{Klaver12}, with a total binding energy in the range
$1.00\pm 0.29$ eV.\footnote{The addition of a further stacking unit,
  to form a pentavacancy cluster, was found to be of similar stability
  to the those considered previously\cite{Klaver12}.} The
corresponding vac$_4$-Y clusters were then investigated by replacing a
central Fe atom with a Y solute. In contrast to the pure Fe case,
where the binding energy is around 0.4 eV higher than for the
trivacancy, the total binding energies for these vac$_4$-Y clusters,
at around 4.9 eV, are lower than the most stable vac$_3$-Y cluster. We
also considered forming vac$_4$-Y clusters by placing a Y solute
within the most open vac$_5$ clusters, which are square-based
pyramidal in form\cite{Klaver12}. While the total binding energy did
increase to 5.06 eV, $\ebind^\mathrm{vac}$ is still negative [see
  \reftab{vacYClusterTab}]. We conclude that the concentration of
vac$_4$-Y clusters will be relatively low in thermal
equilibrium. Finally, we considered the vac$_5$-Y cluster with a Y
atom at the centre of an octahedral hexavacancy\cite{Klaver12}, as
this structure was found to be highly-stable in fcc
Cu\cite{Vineyard1961}, fcc Al\cite{Wang2011} and fct afmD
Fe\cite{Klaver12}.  The total binding energy increases significantly,
which confirms the cluster's stability. Their growth, however, will be
inhibited by the instability of the vac$_4$-Y cluster. Divacancy
absorption by a vac$_3$-Y cluster represents an alternative, and
plausible, formation mechanism, although in this case limited by the
divacancy concentration.

Overall, the presence of Y (and other oversized solutes) in the Fe
matrix provide a high capacity for the trapping of vacancies,
primarily through the formation of highly-stable clusters, such as
vac$_3$-Y and vac$_5$-Y. Not only will this reduce the effective
mobility of vacancies but these clusters should act as natural
recombination sites, reducing the net concentrations of both vacancy
and self-interstitial point defects in the metal. While the oversized
solutes do act as nucleation sites for voids, their net effect will be
to inhibit the growth of large voids and interstitial loops and reduce
the swelling of the material under irradiation. These observations
also suggest the possibility that if any oversized solutes used in the
production of ODS steels, such as Y, Hf and Ti, remain dissolved in
the Fe matrix they would contribute to the observed radiation-damage
resistance of these materials and complement the action of the complex
oxide nanoparticles as point defect sinks and recombination
sites\cite{Brodrick2014,Oka2011}.

\subsection{TM Solute interactions with self-interstitial defects}
\label{siSection}

\begin{figure}[htbp]
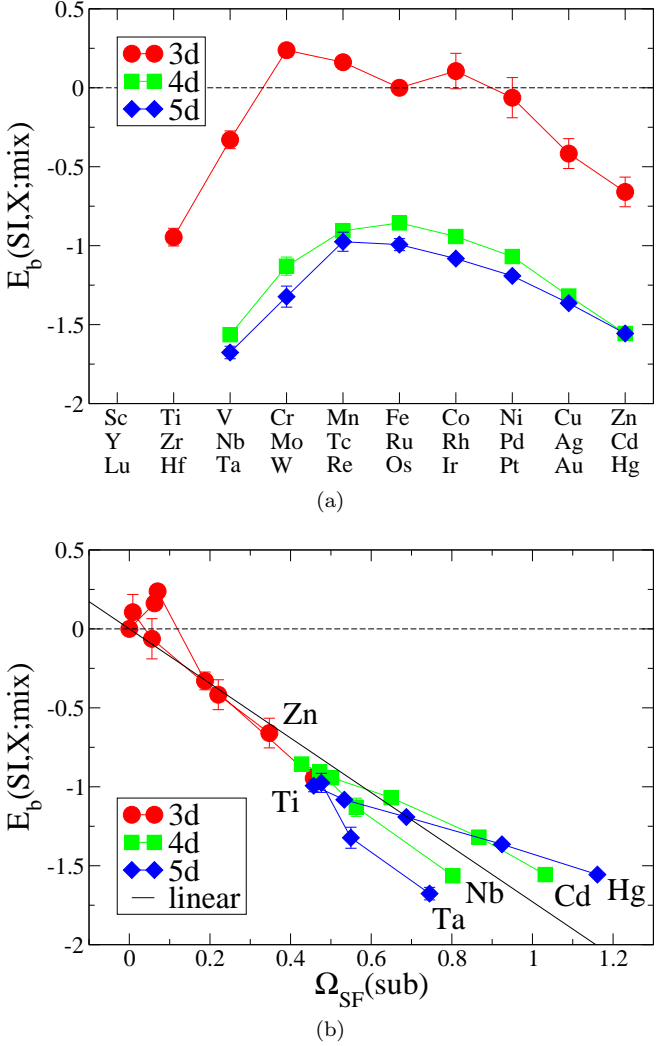

\subfigure[]{\label{EbSISolFeXAcrossFig}\includegraphics[trim = 0mm 0mm 0mm 0.5mm,clip,width=\columnwidth]{Fig12a_Eb-SI-sol-FeX-ave_across_group_vs_nd.eps}}
\vspace{12pt}
\subfigure[]{\label{EbSISolFeXvsSFFig}\includegraphics[trim = 0mm 0mm 0mm 0.5mm,clip,width=\columnwidth]{Fig12b_Eb-SI-sol-FeX-ave_vs_SF-sub.eps}}
\caption{\label{EbSISolFeXFig}SI-solute binding energies for the mixed
  dumbbell configuration, $\ebind(\mathrm{SI,X;mix})$, in eV (a)
  across the TM series and (b) versus the solute size factor,
  $\sfsub$, in fct afmD Fe. The error bars identify the spread in
  binding energies over the two distinct mixed dumbbell configurations
  in the fct afmD structure, with the data point chosen at the centre
  of this range. Note that for Ag, Cd and Hg only one of the mixed
  dumbbells was found to be stable. Figure (b) also shows the result
  of a linear fit to the combined dataset, $\ebind =
  -1.73\ \sizefactor$. The data is given in Appendix \ref{AppB}.}
\end{figure}

In austenite, as in all other fcc metals, the $\langle 001\rangle$
dumbbell is the most stable self-interstitial defect\cite{Klaver12}
and is highly mobile with a migration energy in the range from 0.20 to
0.25 eV\cite{Hepburn13}. The dumbbell produces an anisotropic
distortion of the local lattice, putting the neighbouring atoms under
either compression or tension, which generally leads to repulsion or
attraction to oversized solutes placed in these sites,
respectively\cite{Olsson10}. In this work we have studied the
interactions of TM solutes with the [001] self-interstitial dumbbell
(SI), paying particular attention to those configurations exhibiting
positive binding, where the solutes can act as traps for
self-interstitial defects. We start, however, by considering the
solute binding energies in the mixed dumbbell,
$\ebind(\mathrm{SI,X;mix})$, which is the most compressive solute
environment. The results are shown in \reffig{EbSISolFeXFig}.

The interactions are, generally, repulsive with a strength that
increases with the solute size factor. In fact, the binding energy
data can be successfully modelled as a linear function of the size
factor with a proportionality constant of -1.73 eV, which compares
with a value of -2.03 eV in bcc Fe\cite{Olsson10}. The solute atom in
a mixed dumbbell was also observed to move progressively closer to the
dumbbell lattice site, at the expense of its Fe partner, as the size
factor increased. For the largest solutes, namely Sc, Y, Lu, Zr, Hf,
Ag, Cd and Hg, this tendency resulted in (at least one of) the mixed
dumbbells becoming unstable, with the solute, effectively, occupying
the lattice site and pushing its Fe partner away to form an SI in
either the 2b or 2c configuration [see
  \reffig{DefectSubConfigFig}]. In contrast to these general results,
the magnetic elements Cr, Mn, Co and (to some extent) Ni bind
positively to the mixed dumbbell. The attractive interactions for Cr
and Mn, in particular, stand clearly apart from the general trend with
size factor [see \reffig{EbSISolFeXvsSFFig}]. There is, however, some
consistency in their interactions with point defects, as they are
repelled from the vacancy [see \reffig{EbVacSol1nnFig}], exhibiting
behaviour that would be intuitively expected of undersized solutes,
despite their observed sizes\cite{Klaver12}.

As well as the mixed dumbbell, configurations where the solute
occupies a compressive site at 1nn to the SI [sites 1b and 1c in
  \reffig{DefectSubConfigFig}] are critically important in
interstitial-mediated solute diffusion\cite{Tucker10}. For the 3d
solutes, the trend in binding energy data [see Appendix \ref{AppB}]
follows a very similar pattern to that for the mixed dumbbell. Once
again, Cr, Mn, Co and (to some extent) Ni exhibit positive binding
while the oversized solutes are repelled, although to a much lesser
extent than from the mixed dumbbell. It is interesting to note that V
is positively bound to the SI in the 1nn compressive sites, despite
being repelled from the mixed dumbbell. We conclude that
interstitial-mediated diffusion is only likely to be important for the
magnetic solutes, with the effect being most pronounced for Cr and Mn.

\begin{figure}[htbp]
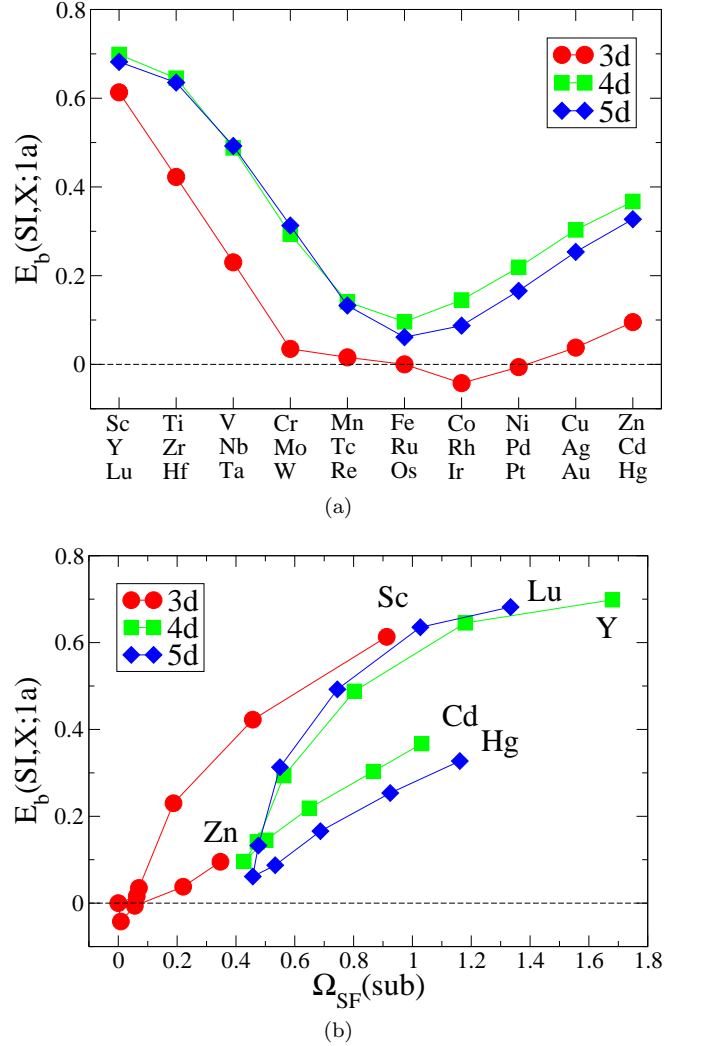

\subfigure[]{\label{EbSISol1nnTensAcrossFig}\includegraphics[trim = 0mm 0mm 0mm 0.5mm,clip,width=\columnwidth]{Fig13a_Eb-SI-sol-1nn-tens_across_group_vs_nd.eps}}
\vspace{12pt}
\subfigure[]{\label{EbSISol1nnTensvsSFFig}\includegraphics[trim = 0mm 0mm 0mm 0.5mm,clip,width=\columnwidth]{Fig13b_Eb-SI-sol-1nn-tens_vs_SF-sub.eps}}
\caption{\label{EbSISol1nnTensFig}SI-solute binding energies for the
  1nn tensile configuration, $\ebind(\textrm{SI,X;1a})$, in eV
  (a) across the TM series and (b) versus the solute size factor,
  $\sfsub$, in fct afmD Fe. This configuration is uniquely defined in
  the fct afmD structure, with the solute at site 1a in
  \reffig{DefectSubConfigFig}. The data is given in Appendix
  \ref{AppB}.}
\end{figure}

In contrast to the two cases above, we observed, almost exclusively,
attractive interactions for solutes in the 1nn tensile site near an SI
[site 1a in \reffig{DefectSubConfigFig}]. The binding energies,
$\ebind(\textrm{SI,X;1a})$, in \reffig{EbSISol1nnTensFig} exhibit
clear trends across the TM series and the data clearly differentiates
between the 3d and 4d/5d solutes. The strength of binding does
increase with the solute size factor but the early and late TMs follow
quite distinct trends [\reffig{EbSISol1nnTensvsSFFig}], as observed
for other quantities here and in bcc Fe\cite{Olsson10}. The binding
energies for the late TM solutes are, approximately, proportional to
their size factors with a proportionality coefficient of around 0.3 eV
and while the binding energies for the early TM solutes do increase at
a greater (non-linear) rate, the data appears to saturate for
$\sizefactor > 1$.

\begin{figure}[htbp]
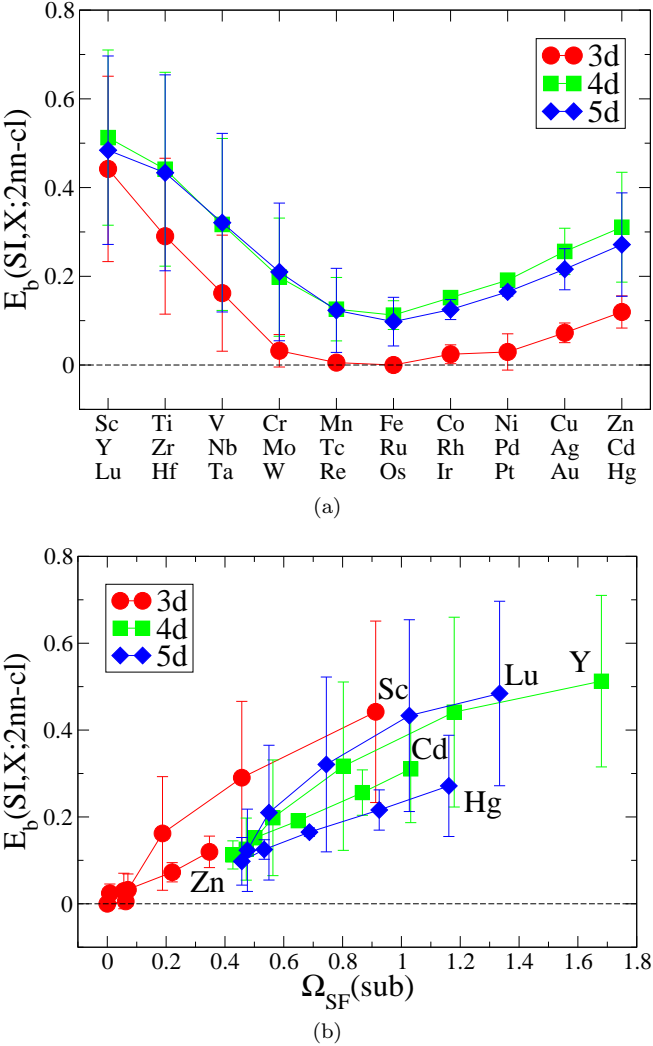

\subfigure[]{\label{EbSISol2nnCollinearAcrossFig}\includegraphics[trim = 0mm 0mm 0mm 0.5mm,clip,width=\columnwidth]{Fig14a_Eb-SI-sol-2nn-collinear-ave_across_group_vs_nd.eps}}
\vspace{12pt}
\subfigure[]{\label{EbSISol2nnCollinearvsSFFig}\includegraphics[trim = 0mm 0mm 0mm 0.5mm,clip,width=\columnwidth]{Fig14b_Eb-SI-sol-2nn-collinear-ave_vs_SF-sub.eps}}
\caption{\label{EbSISol2nnCollinearFig}SI-solute binding energies,
  $\ebind(\textrm{SI,X;2nn-cl})$, in eV for solutes at 2nn and in
  sites collinear with the [001] self-interstitial defect (a) across
  the TM series and (b) versus the solute size factor, $\sfsub$, in
  fct afmD Fe. The error bars identify the spread in binding energies
  over the two distinct sites, namely 2b and 2c in
  \reffig{DefectSubConfigFig}, with the data point chosen at the
  centre of this range. The data is given in Appendix \ref{AppB}.}
\end{figure}

Positive binding energies of up to the same magnitude and following
very similar (average) trends [see \reffig{EbSISol2nnCollinearFig}]
were also observed for solutes in the 2nn sites collinear with the
[001] dumbbell [sites 2b and 2c in \reffig{DefectSubConfigFig}]. While
the large spread in the data does preclude a detailed analysis, the
binding energy clearly increases with solute size factor. Calculations
for the 3d solutes in the 2a site [see \reffig{DefectSubConfigFig}]
found only weak binding [see Appendix \ref{AppB}], that was positively
correlated to the solute size factor.

Overall, we have demonstrated that oversized TM solutes can act as
strong trapping sites for SI defects and that this effect increases
with the solute size factor. Their addition to austenitic steels
should, therefore, not only act to reduce the effective mobility of SI
defects but lead to enhanced recombination rates and a reduction in
the net defect concentrations under irradiation. The data also
suggests that oversized TM solutes will act as nucleation sites for
self-interstitial clusters.

\section{Conclusions}

In this work we have extended the theoretical database of atomic-level
properties of steels by performing a comprehensive set of first
principles electronic structure calculations to study transition metal
solute properties in austenite and their interactions with point
defects.

We have found clear trends in the properties of substitutional TM
solutes in the defect-free lattice, their binding energies to point
defects and in quantities relevant to vacancy-mediated solute
diffusion across the TM series, as a function of the local d band
occupancy of the solute atoms. The interaction data between TM solutes
and point defects are highly correlated to the solute size factors in
a way that is consistent with arguments based on elasticity and local
strain field effects. Functional dependencies can generally be used to
model these relationships, although in some cases the early and late
TM solutes show quite distinct behaviour, as observed in bcc
Fe\cite{Olsson10}. Throughout this work we have observed high levels
of consistency and strong correlation between results in fct afmD Fe
and bcc fm Fe\cite{Olsson10}, which adds to similar observations made
previously\cite{Klaver12,Hepburn13}. We would expect this
insensitivity to the crystal structure to extend to other solvent
metals.

We have shown that oversized TM solutes act as strong traps for both
vacancy and self-interstitial defects, with a strength that increases
with the solute size factor. Furthermore, we have shown, using Y as a
representative, that oversized solutes act as strong traps for
additional vacancies, forming close-packed vacancy clusters around a
central solute. The vac$_3$-X and vac$_5$-X clusters were found to
form particularly stable configurations. Our previous
analysis\cite{Klaver12} suggests that highly-stable clusters of
aligned self-interstitial dumbbells should form around a single solute
atom. This high trapping capacity should result in a significant
lowering of defect mobility and reduction in the net concentration of
point defects in the matrix, both by enhancing defect recombination
and by providing nucleation sites for the formation of secondary
defects, such as proto-voids\cite{Kato1991} and interstitial
loops. Overall, these observations provide a strong foundation for the
suggestion by Kato {\it et al.}\cite{Kato1991,Kato1992}, that point
defect trapping at oversized TM solutes underlies their experimental
observations of reduced swelling\cite{Kato1991} and a decrease in
RIS\cite{Kato1992} in 316L austenitic steel doped with small
concentrations of solutes. The same conclusions also apply to ODS
steels where any oversized solutes, such as Y, Hf and Ti, remaining
dissolved in the Fe matrix after manufacture would contribute to the
radiation-damage resistance provided by the oxide
nanoparticles\cite{Brodrick2014,Oka2011}.

We have extended our previous analysis\cite{Klaver12} of
vacancy-mediated solute diffusion to cover Cr, Mn, Co, Ni and Cu. We
find that Ni and Co diffuse at similar rates below that of Fe and will
diffuse with the vacancy flux by the vacancy drag mechanism below a
critical temperature, which for Co is $400\pm 50$ K.  In contrast,
both Cr and Cu diffuse more quickly than Fe and Mn at an intermediate
rate against the vacancy flux. We infer that the concentrations of Co
and Ni will be enhanced and those of Cr and Cu depleted at defect
sinks.

We have demonstrated a reduction in the migration barrier for
vacancy-solute exchange at 1nn ($\omega_2$ jumps) as the solute size
factor increases. For sufficiently large solutes, namely Sc, Y, Zr, Lu
and Hf, the barrier ceases to exist and the solute, X, stably binds to
the vacancy at a position half way between the two lattice sites to
form an SCD defect. This is the transition state configuration for the
smaller solutes. For those solutes forming a stable SCD, or for those
where the $\omega_2$ jump barrier is below the thermal energy
$k_\mathrm{B}T$, namely Ti, Nb and Ta, vacancy-mediated solute
diffusion is dominated by the $\omega_3$ dissociation/reassociation
mechanism identified in this work. The activation energy for this
process is lower than that for self-diffusion, which is in contrast to
the, often assumed, immobility of such large solutes. This important
result should be taken account of in future studies of the nucleation
and growth of complex oxide nanoparticles in ODS steels.

Interstitial-mediated solute diffusion will be energetically
disfavoured in proportion to the solute size factor for all solutes
except Cr, Mn, Co and (to a lesser extent) Ni, where, magnetic effects
lead to favourable interactions with the self-interstitial
defect. Even for these solutes, the relative contribution compared to
vacancy-mediated diffusion will depend critically on the
concentrations of the respective defects in the matrix and a
definitive study is well beyond the scope of this
work\cite{Tucker10}.

Finally, we note that since the majority of our conclusions are based
on solute size factor effects they should generalise to other solvent
metals and to concentrated austenitic steels in particular.

\section{Acknowledgements}

This work was part sponsored through the EU-FP7 PERFORM-60 project,
the G8 funded NuFUSE project and EPSRC through the UKCP collaboration.

\appendix

\section{Elemental data and ground state crystal structure calculations}
\label{AppA}

We performed a set of high-precision ab initio calculations for the
ground state (0K) crystalline and magnetic structures of all the
transition metals (TMs), primarily for use as reference states in the
determination of formation energies for the TM solute calculations
presented in this work. Aside from Mn and Fe, these crystal structures
are either hexagonal close-packed (hcp), body-centred cubic (bcc),
face-centred cubic (fcc) or rhombohedral (rho) and the magnetic
structures either non-magnetic (nm), ferromagnetic (fm) or
antiferromagnetic (afm). The ground state crystallographic parameters
were determined by full relaxation of the unit cell and atomic
positions. To ensure that the unit cell stress tensor and, therefore,
lattice parameters were determined accurately we used a plane-wave
energy cutoff of 550 eV, a high-density Monkhorst-Pack k-point grid to
sample the Brillouin zone [see \reftab{elementsTab}] and an energy
tolerance of $10^{-8}$ eV to converged the electronic ground
state. For the structural relaxations, forces were converged to less
than $10^{-4}$ eV/$\angs$ and the cell stress to less than $5\times
10^{-6}$ eV/$\angs^3$ (0.008 kB). Detailed calculations showed that
these settings were sufficient to converge the energy to better than
0.5 meV/atom, the pressure to $5\times 10^{-4}$ eV/$\angs^3$ (0.8 kB)
and local magnetic moments to $10^{-3}\ \magneton$, resulting in
errors to the lattice parameters of much less than 0.001 $\angs$. The
results are given in \reftab{elementsTab} for all elements except Mn,
which we now discuss in more detail.

\begin{table*}[htbp]
\begin{ruledtabular}
\begin{tabular}{llllll}
Element & $N_\mathrm{val}$ : Config. & $r_\mathrm{wigs}$ & k-points & Crystal structure and parameters & $E_\mathrm{coh}$ \\ 
\hline 
Sc & 11 : 3s$^2$3p$^6$3d$^1$4s$^2$ & 1.429 & 16x16x12 & hcp, nm, $a=3.314\ \angs$, $c=5.144\ \angs$ & 3.90 \\
Ti & 10 : 3p$^6$3d$^2$4s$^2$ & 1.323 & 16x16x12 & hcp, nm, $a=2.932\ \angs$, $c=4.642\ \angs$ & 4.85 \\
V & 11 : 3p$^6$3d$^3$4s$^2$ & 1.217 & 20x20x20 & bcc, nm, $a=2.996\ \angs$ & 5.31 \\
Cr & 6 : 3d$^5$4s$^1$ & 1.323 & 20x20x20 & bcc, afm, $a=2.849\ \angs$, $|\mu|=0.92\ \magneton$ & 4.10 \\
Mn & 7 : 3d$^5$4s$^2$ & 1.323 & 6x6x6 & $\alpha$-Mn [see \reftab{manganeseTab}] & 2.92 \\
Fe & 8 : 3d$^6$4s$^2$ & 1.302 & 20x20x20 & bcc, fm, $a=2.832\ \angs$, $\mu=2.20\ \magneton$ & 4.28 \\
 & & & 16x16x8 & fct, afmD, $a=3.447\ \angs$, $c=3.750\ \angs$, $|\mu|=1.99\ \magneton$ & 4.20 \\
 & & & 16x16x16 & fcc, nm, $a=3.447\ \angs$ & 4.06 \\
Co & 9 : 3d$^7$4s$^2$ & 1.302 & 18x18x10 & hcp, fm, $a=2.495\ \angs$, $c=4.028\ \angs$, $\mu=1.62\ \magneton$ & 4.39 \\
Ni & 10 : 3d$^8$4s$^2$ & 1.286 & 18x18x18 & fcc, fm, $a=3.522\ \angs$, $\mu=0.63\ \magneton$ & 4.44 \\
Cu & 11 : 3d$^{10}$4s$^1$ & 1.312 & 18x18x18 & fcc, nm, $a=3.636\ \angs$ & 3.49 \\
Zn & 12 : 3d$^{10}$4s$^2$ & 1.270 & 20x20x16 & hcp, nm $a=2.643\ \angs$, $c=5.080\ \angs$ & 1.35 \\
\hline
Y & 11 : 4s$^2$4p$^6$4d$^1$5s$^2$ & 1.815 & 16x16x12 & hcp, nm, $a=3.649\ \angs$, $c=5.661\ \angs$ & 4.37 \\
Zr & 12 : 4s$^2$4p$^6$4d$^2$5s$^2$ & 1.625 & 16x16x12 & hcp, nm, $a=3.232\ \angs$, $c=5.180\ \angs$ & 6.25 \\
Nb & 11 : 4p$^6$4d$^4$5s$^1$ & 1.503 & 20x20x20 & bcc, nm, $a=3.323\ \angs$ & 7.57 \\
Mo & 12 : 4p$^6$4d$^5$5s$^1$ & 1.455 & 20x20x20 & bcc, nm, $a=3.172\ \angs$ & 6.82 \\
Tc & 13 : 4p$^6$4d$^6$5s$^1$ & 1.423 & 20x20x16 & hcp, nm, $a=2.764\ \angs$, $c=4.420\ \angs$ & 6.85 \\
Ru & 8 : 4d$^7$5s$^1$ & 1.402 & 20x20x16 & hcp, nm, $a=2.729\ \angs$, $c=4.304\ \angs$ & 6.74 \\
Rh & 9 : 4d$^8$5s$^1$ & 1.402 & 18x18x18 & fcc, nm, $a=3.844\ \angs$ & 5.75 \\
Pd & 10 : 4d$^{10}$5s$^0$ & 1.434 & 18x18x18 & fcc, nm, $a=3.956\ \angs$ & 3.89 \\
Ag & 11 : 4d$^{10}$5s$^1$ & 1.503 & 18x18x18 & fcc, nm, $a=4.157\ \angs$ & 2.95 \\
Cd & 12 : 4d$^{10}$5s$^2$ & 1.577 & 20x20x16 & hcp, nm, $a=3.023\ \angs$, $c=5.798\ \angs$ & 1.16 \\
\hline
Lu & 9 : 5p$^6$5d$^1$6s$^2$ & 1.588 & 16x16x12 & hcp, nm, $a=3.514\ \angs$, $c=5.460\ \angs$ & 4.43 \\
Hf & 10 : 5p$^6$5d$^2$6s$^2$ & 1.614 & 20x20x16 & hcp, nm, $a=3.199\ \angs$, $c=5.054\ \angs$ & 6.44 \\
Ta & 11 : 5p$^6$5d$^3$6s$^2$ & 1.503 & 20x20x20 & bcc, nm, $a=3.320\ \angs$ & 8.10 \\
W & 12 : 5p$^6$5d$^4$6s$^2$ & 1.455 & 20x20x20 & bcc, nm, $a=3.190\ \angs$ & 8.90 \\
Re & 7 : 5d$^5$6s$^2$ & 1.455 & 18x18x14 & hcp, nm, $a=2.779\ \angs$, $c=4.485\ \angs$ & 8.03 \\
Os & 14 : 5p$^6$5d$^6$6s$^2$ & 1.413 & 20x20x16 & hcp, nm, $a=2.761\ \angs$, $c=4.357\ \angs$ & 8.17 \\
Ir & 9 : 5d$^9$6s$^0$ & 1.423 & 18x18x18 & fcc, nm, $a=3.882\ \angs$ & 6.94 \\
Pt & 10 : 5d$^9$6s$^1$ & 1.455 & 18x18x18 &  fcc, nm, $a=3.985\ \angs$ & 5.84 \\
Au & 11 : 5d$^{10}$6s$^1$ & 1.503 & 18x18x18 & fcc, nm, $a=4.173\ \angs$ & 3.81 \\
Hg & 12 : 5d$^{10}$6s$^2$ & 1.614 & 26x26x26 & rho, nm, $a=3.101\ \angs$, $\gamma=84.4^\circ$ & 0.67 \\
\end{tabular}
\end{ruledtabular}
\caption{\label{elementsTab} Calculation details and ground state
  properties for the transition metals elements considered in this
  work. In particular we give the number of valence electrons in the
  PAW potential, $N_\mathrm{val}$, and the atomic valence electron
  configuration\cite{Kittel}, the atomic radius used for calculating
  local magnetic moments, $r_\mathrm{wigs}$, in $\angs$, the
  dimensions of the k-point grid used to sample the Brillouin zone,
  the equilibrium crystal and magnetic structure parameters for the
  conventional unit cell (at T=0K) and the experimental cohesive
  energy per atom at T=0K [from Kittel\cite{Kittel}, p.50],
  $E_\mathrm{coh}$, in eV. The values of $E_\mathrm{coh}$ for the fct
  afmD and fcc nm states of Fe were estimated using the ab initio
  energy differences to the bcc fm ground state of 0.077 and 0.216
  eV/atom, respectively\cite{Klaver12}, and experimental ground state
  cohesive energy. }
\end{table*}

The crystalline structure of Mn differs distinctly from the other
transition metals. Under standard conditions of temperature and
pressure the most stable polymorph, $\alpha$-Mn, is paramagnetic
(para) with a 58 atom body-centred cubic unit cell with space group
$\mathrm{T}^3_\mathrm{d}-\mathrm{I}\bar43\mathrm{m}$ (number 217), as
first resolved by Bradley and Thewlis\cite{Bradley1927}. They found a
lattice parameter, $a = 8.894\ \angs$, and four crystallographically
distinct sets of atomic positions. Using the nomenclature of Hobbs
{\it et al.}\cite{Hobbs01,Hobbs03} their number, Wyckoff positions and
internal coordinates relative to
$[(0,0,0),(\frac{1}{2},\frac{1}{2},\frac{1}{2})]$ are as follows: 2
type-I atoms at (a), $[(0,0,0)]$; 8 type-II atoms at (c),
$[(x,x,x),(x,-x,-x),(-x,x,-x),(-x,-x,x)]$ and two sets of 24 atoms,
type-III and type-IV, at (g),
$[(x,x,z),(x,-x,-z),(-x,x,-z),(-x,-x,z)]+$ cyclic permutations.  A
more recent and accurate study by Yamada {\it et al.} used single
crystal measurements to extrapolate the crystallographic parameters of
para $\alpha$-Mn to 0 K\cite{Yamada1970}. The results are summarised
in \reftab{manganeseTab}.

\begin{table*}[htbp]
\begin{ruledtabular}
\begin{tabular}{cccccccc}
\multirow{2}{*}{Authors} & Bradley \&                   & Yamada                        & Lawson                        & Hobbs                      & Hobbs                      & This & This \\ 
                         & Thewlis\cite{Bradley1927}    & {\it et al.}\cite{Yamada1970} & {\it et al.}\cite{Lawson1994} & {\it et al.}\cite{Hobbs03} & {\it et al.}\cite{Hobbs03} & work & work \\
Magnetism & para & para & afm & nm & afm & nm & afm \\
\hline 
$a$ &  \multirow{2}{*}{8.894} & \multirow{2}{*}{8.865} & 8.877 & \multirow{2}{*}{8.532} & 8.669 & \multirow{2}{*}{8.546} & \multirow{2}{*}{8.636} \\
$c$ & & & 8.873 & & 8.668 & & \\
& & & & & & & \\
$V_\mathrm{atom}$ & 12.13 & 12.01 & 12.06 & 10.71 & 11.23 & 10.76 & 11.10 \\
& & & & & & & \\
$x(\mathrm{II})$ & \multirow{2}{*}{$0.31_7$} & \multirow{2}{*}{0.317} & 0.3192(2) & \multirow{2}{*}{0.318} & 0.320 & \multirow{2}{*}{0.318} & \multirow{2}{*}{0.319} \\
$z(\mathrm{II})$ & & & 0.3173(3) & & 0.319 & & \\
& & & & & & & \\
$x(\mathrm{IIIa})$ & \multirow{3}{*}{$0.35_6$} & \multirow{3}{*}{0.357} & 0.3621(1) & \multirow{3}{*}{0.356} & 0.355 & \multirow{3}{*}{0.356} & \multirow{3}{*}{0.356} \\
$x(\mathrm{IIIb})$ & & & 0.3533(2) & & 0.355 & & \\
$z(\mathrm{IIIb})$ & & & 0.3559(2) & & 0.354 & & \\
& & & & & & & \\
$z(\mathrm{IIIa})$ & \multirow{2}{*}{$0.04_2$} & \multirow{2}{*}{0.034} & 0.0408(2) & \multirow{2}{*}{0.037} & 0.032 & \multirow{2}{*}{0.037} & \multirow{2}{*}{0.035} \\
$y(\mathrm{IIIb})$ & & & 0.0333(1) & & 0.033 & & \\
& & & & & & & \\
$x(\mathrm{IVa})$ & \multirow{3}{*}{$0.08_9$} & \multirow{3}{*}{0.089} & 0.0921(2) & \multirow{3}{*}{0.088} & 0.088 & \multirow{3}{*}{0.088} & \multirow{3}{*}{0.088} \\
$x(\mathrm{IVb})$ & & & 0.0895(2) & & 0.088 & & \\
$z(\mathrm{IVb})$ & & & 0.0894(2) & & 0.087 & & \\
& & & & & & & \\
$z(\mathrm{IVa})$ & \multirow{2}{*}{$0.27_8$} & \multirow{2}{*}{0.282} & 0.2790(3) & \multirow{2}{*}{0.281} & 0.283 & \multirow{2}{*}{0.281} & \multirow{2}{*}{0.283} \\
$y(\mathrm{IVb})$ & & & 0.2850(1) & & 0.283 & & \\
& & & & & & & \\
$\mu(\mathrm{I})$    & --- & --- &  2.83(13) & --- &  2.79 & --- & 2.86 \\
$\mu(\mathrm{II})$   & --- & --- &  1.83(06) & --- &  2.22 & --- & 2.31 \\
$\mu(\mathrm{IIIa})$ & --- & --- &  0.74(14) & --- & -1.11 & --- & -1.23 \\
$\mu(\mathrm{IIIb})$ & --- & --- & -0.48(11) & --- & -1.10 & --- & -1.23 \\
$\mu(\mathrm{IVa})$  & --- & --- & -0.59(10) & --- & 0.0 & --- & $|\mu|$ \\
$\mu(\mathrm{IVb})$  & --- & --- &  0.66(07) & --- & 0.0 & --- & $< 0.03$ \\
\end{tabular}
\end{ruledtabular}
\caption{\label{manganeseTab} Crystallographic parameters for
  $\alpha$-Mn. The lattice parameters, $a$ and $c$, are in $\angs$,
  the atomic volume, $V_\mathrm{atom}$, in $\angs^3$ and the other
  internal parameters are dimensionless. The results of Yamada {\it et
    al.}\cite{Yamada1970} are for para $\alpha$-Mn extrapolated to 0
  K. The results of Lawson {\it et al.}\cite{Lawson1994} were measured
  at 15 K. Magnetic moments, $\mu$, are given in $\magneton$ for the
  distinct atomic types centred on (0,0,0), with the moments around
  $(\frac{1}{2},\frac{1}{2},\frac{1}{2})$ anti-parallel to these. For
  the non-collinear structure of Lawson {\it et al.}\cite{Lawson1994}
  the magnitudes of the moments are given and the sign indicates the
  moment direction when projected onto the MnI moments about
  (0,0,0). It should be noted that the relative orientations of the
  moments from Hobbs {\it et al.}\cite{Hobbs03} were determined from
  figures in that work given the lack of clarity in their
  specification in the text and tables.}
\end{table*}

Low-temperature neutron diffraction studies by Shull and
Wilkinson\cite{Shull1953} found that $\alpha$-Mn is afm below a N\'eel
temperature of 95 K. Further studies to resolve the magnetic
structure\cite{Kasper1956,Oberteuffer1968,Kunitomi1969,Yamada1970,Lawson1994}
were complicated by the need to use theoretical models to analyse and
interpret the diffraction data, resulting in a number of both
collinear and non-collinear magnetic structures exhibiting a whole
range of magnetic moments. Kunitomi {\it et al.}\cite{Kunitomi1969}
showed that a non-collinear model was necessary to reproduce the
experimental results and the magnetic structure then resolved by
Yamada {\it et al.}\cite{Yamada1970} following a group-theoretical
approach\cite{Yamada1970b}, although with some remaining variability
in the moments depending on the exact details of the model used. More
recent work by Lawson {\it et al.}\cite{Lawson1994} used a Shubnikov
(magnetic space) group-based analysis, yielding an anti-body-centred
tetragonal magnetic structure, equivalent to Yamada {\it et
  al.}\cite{Yamada1970}. Furthermore, they were able to determine that
the implied body-centred and weakly tetragonal crystal structure
belongs to space group $\mathrm{I}\bar42\mathrm{m}$ (number 121), with
the four distinct sets of atoms in the paramagnetic case now split
into six: The 2 type-I atoms are unchanged, the 8 type-II atoms now
take Wyckoff position (i), $[(x,x,z),(x,-x,-z),(-x,x,-z),(-x,-x,z)]$
and the 24 type-III and type-IV atoms now split into two distinct
subsets with 8 atoms (IIIa/IVa) at position (i) and 16 (IIIb/IVb) at
(j), $[(x,y,z)$, $(x,-y,-z)$, $(-x,y,-z)$, $(-x,-y,z)$, $(y,x,z)$,
  $(y,-x,-z)$, $(-y,x,-z)$, $(-y,-x,z)]$. Determinations of the
crystallographic parameters at a number of temperatures from 305 to 15
K\cite{Lawson1994} clearly shows the onset of the magnetic transition
with its coupled tetragonal distortion of the lattice and the
splitting of the internal coordinates below the N\'eel temperature so
that $x(\mathrm{II}) \ne z(\mathrm{II})$, $x(\mathrm{IIIa}),
x(\mathrm{IIIb}), z(\mathrm{IIIb}) \ne x(\mathrm{III})$ and
$z(\mathrm{IIIa}), y(\mathrm{IIIb}) \ne z(\mathrm{III})$, with
equivalent results for the type-IV atoms. Along with Bradley and
Thewlis\cite{Bradley1927} they also make the interesting point that
the complexity of the $\alpha$-Mn structure (as compared to the other
TMs) can be understood once viewed as a self-intermetallic compound
between Mn atoms in crystallographically distinct sites with distinct
electronic/magnetic configurations and, therefore, different atomic
sizes. The results of Lawson {\it et al.}\cite{Lawson1994} are
summarised in \reftab{manganeseTab}.

Theoretical attempts to model $\alpha$-Mn culminate in a comprehensive
ab initio study by Hobbs {\it et al.}\cite{Hobbs01,Hobbs03}, who also
provide an excellent summary and discussion of the preceding
theoretical and experimental work on Mn. The other polymorphs of Mn
are considered in related work\cite{Hobbs01,Hobbs03II,Hafner05}. Their
study covers the nm state and both collinear and non-collinear afm
magnetic states of $\alpha$-Mn over a range of atomic volumes. For the
nm state they find a low equilibrium atomic volume of 10.71 $\angs^3$
($a=8.532\ \angs$). The equilibrium afm state lies around 0.025
eV/atom lower than the nm state (as determined from their energy vs
volume curves) at an atomic volume of 11.23 $\angs^3$ and exhibits a
collinear magnetic structure with only marginal evidence of any
tetragonal distortion [see \reftab{manganeseTab}]. It is only above
the experimental volume (~12 $\angs^3$/atom) that any appreciable
non-collinearity in the magnetic structure and tetragonal lattice
distortion is observed, which they suggest is closely related to the
critical development of non-zero moments on MnIV atoms.

The results of our own calculations are summarised in
\reftab{manganeseTab}. We find that the nm state of $\alpha$-Mn has an
equilibrium volume of 10.76 $\angs^3$ ($a=8.546\ \angs$), in good
agreement with Hobbs {\it et al.}\cite{Hobbs01,Hobbs03}. Our use of a
finer $6^3$ k-point grid may explain the slight discrepancy. It is
often said that Mn would resort to an hcp structure, like the other
group VII TMs Tc and Re, in the absence of magnetism. We, however,
find the surprising result that the equilibrium nm hcp structure
($a=2.478\ \angs$, $c=4.004\ \angs$) lies 45 meV/atom above nm
$\alpha$-Mn. This also indicates that the primary mechanism driving
the formation of the complex $\alpha$-Mn structure is not magnetic in
origin.

Determination of the afm structure was significantly more complex. For
the magnetic structure we initialised the moments on MnIV atoms to
zero, following Hobbs {\it et al.}\cite{Hobbs03}. For consistency with
the experimental and theoretical results in the literature we take the
moments on atoms of the same type to be equal in magnitude but with
anti-parallel orientations about (0,0,0) and
$(\frac{1}{2},\frac{1}{2},\frac{1}{2})$ (to produce the afm
structure). With these assumptions there are still 16 distinct
relative orientations of moments between the different atomic types
for the tetragonal structure. Calculations were initialised in all of
these distinct magnetic states with either cubic\cite{Gazzara1966} or
tetragonal\cite{Lawson1994,Hobbs03} lattice parameters. Despite many
distinct magnetic states being initially stable only one stable afm
structure was found after full relaxation [see \reftab{manganeseTab}]. 

We found a cubic afm structure with an atomic volume of 11.10
$\angs^3$ ($a=8.636\ \angs$), which is 8.0\% (2.7\%) lower than
experiment\cite{Lawson1994}, although this is typical of GGA
calculations on afm systems\cite{Hobbs03}. We found no evidence of a
stable tetragonally distorted lattice, unlike Hobbs {\it et
  al.}\cite{Hobbs01,Hobbs03} although their calculations only show a
very marginal effect. The energy difference between the nm and afm
states of $\alpha$-Mn, that we measure to be 28 meV/atom, does,
interestingly, agree well with Hobbs {\it et al.}. The internal
coordinates show a high degree of consistency both with the nm state
from this work and with other theoretical\cite{Hobbs01,Hobbs03} and
experimental\cite{Bradley1927,Yamada1970,Lawson1994,Gazzara1966} work,
although this is, perhaps, not surprising given their relative
invariance as a function of temperature above and below the magnetic
transition\cite{Lawson1994}.  For the magnetic structure we find large
moments on MnI and MnII atoms, that agree qualitatively with the
(near)-collinear moments found in other work\cite{Hobbs03,Lawson1994},
and smaller moments on MnIII and MnIV atoms, consistent with the
majority of previous studies [see Hobbs {\it et al.}\cite{Hobbs03} and
  references therein]. While the MnIII moments are similar in
magnitude to those from experiment\cite{Lawson1994} we found that our
calculations did not differentiate between MnIIIa and MnIIIb atoms,
despite initialising their positions consistent with a tetragonal
structure\cite{Hobbs03,Lawson1994} and their moments to be either
parallel or anti-parallel and with different magnitudes. Along with
Hobbs {\it et al.}\cite{Hobbs03} we also found near-zero equilibrium
moments on MnIV atoms, in contrast with experiment\cite{Lawson1994}.
Given that Hobbs {\it et al.}\cite{Hobbs01,Hobbs03} report the
generation of non-collinearity in MnIII and MnIV moments as well as
non-zero MnIV moments at volumes exceeding equilibrium, it is
certainly plausible that the failure of theory to produce the correct
magnetic state at equilibrium is closely related to its
underestimation of the atomic volume. Overall, we conclude that the
afm state we have found is the best-possible reproduction of the
ground-state structure for $\alpha$-Mn within the particular
theoretical framework used in this work.

\section{TM solute data}
\label{AppB}

In this appendix, we present the data from the large supercell
calculations used in this work. The data is given at the precision of
the VASP output for reproducibility and further use and should not be
taken to indicate the accuracy of the results. Substitutional TM
solute properties in fct afmD and bcc fm Fe are given in
\reftab{soluteDataTab}. Vacancy-solute binding energies at 1nn
separation and vacancy migration energies for the five-frequency model
jumps in fct afmD Fe are given in \reftab{vacDataTab}. Binding
energies between TM solutes and an [001] self-interstitial dumbbell at
up to 2nn separation in fct afmD Fe are given in
\reftab{siDataTab}. Vacancy-Y binding energies at 2nn, 3nn and 4nn
separations in fct afmD Fe are given in \reftab{EbYDataTab}. Migration
energies for $\omega_3$ vacancy jumps near a Y solute in fct afmD Fe
are given in \reftab{EmYDataTab}.

\begin{table*}[htbp]
\begin{ruledtabular}
\begin{tabular}{lcccccc}
& \multicolumn{3}{c}{fct afmD Fe} & \multicolumn{3}{c}{bcc fm Fe} \\
& $\eformsub$ & $\mu(\mathrm{sub})$ & $\sfsub$ & $\eformsub$ & $\mu(\mathrm{sub})$ & $\sfsub$ \\
\hline
Sc &  0.423638 & -0.099 & 0.913 &  0.315274 & -0.394 & 0.665 \\
Ti & -0.376736 & -0.144 & 0.457 & -0.805544 & -0.757 & 0.381 \\
V  & -0.144885 & -0.070 & 0.188 & --- & --- & --- \\
Cr &  0.271619 &  0.847 & 0.070 & --- & --- & --- \\
Mn &  0.064990 &  1.999 & 0.063 & --- & --- & --- \\
Co &  0.179164 &  0.978 & 0.009 & --- & --- & --- \\
Ni &  0.087110 &  0.039 & 0.056 & --- & --- & --- \\
Cu &  0.511519 & -0.007 & 0.221 &  0.752995 &  0.111 & 0.218 \\
Zn &  0.207554 & -0.013 & 0.347 &  0.326639 & -0.081 & 0.342 \\
\hline
Y  & 1.994622 & -0.084 & 1.680 & 2.094273 & -0.279 & 1.310 \\
Zr & 0.600812 & -0.098 & 1.180 & 0.377658 & -0.467 & 1.015 \\
Nb & 0.378045 & -0.076 & 0.803 & --- & --- & --- \\
Mo & 0.472454 &  0.068 & 0.563 & --- & --- & --- \\
Tc & 0.258085 &  0.238 & 0.472 & --- & --- & --- \\
Ru & 0.265435 &  0.295 & 0.427 & --- & --- & --- \\
Rh & 0.081337 &  0.158 & 0.502 & --- & --- & --- \\
Pd & 0.490826 &  0.017 & 0.649 & --- & --- & --- \\
Ag & 1.756191 & -0.009 & 0.867 & 1.914812 &  0.100 & 0.937 \\
Cd & 1.746557 & -0.012 & 1.032 & 1.883467 & -0.064 & 0.951 \\
\hline
Lu &  1.197321 & -0.109 & 1.334 &  1.233167 & -0.372 & 1.035 \\
Hf &  0.235090 & -0.099 & 1.027 & -0.016113 & -0.468 & 0.891 \\
Ta &  0.128539 & -0.068 & 0.745 & --- & --- & --- \\
W  &  0.457315 &  0.005 & 0.550 & --- & --- & --- \\
Re &  0.243007 &  0.136 & 0.476 & --- & --- & --- \\
Os &  0.233483 &  0.217 & 0.457 & --- & --- & --- \\
Ir & -0.169113 &  0.170 & 0.533 & --- & --- & --- \\
Pt & -0.105044 &  0.044 & 0.687 & --- & --- & --- \\
Au &  1.072340 & -0.006 & 0.924 &  1.069742 &  0.171 & 1.073 \\
Hg &  2.053529 & -0.012 & 1.161 &  2.157507 & -0.031 & 1.197 \\
\end{tabular}
\end{ruledtabular}
\caption{\label{soluteDataTab} The formation energy, $\eformsub$, in
  eV, magnetic moment (in an up-spin magnetic plane for fct afmD Fe),
  $\mu(\mathrm{sub})$, in $\magneton$ and size factor, $\sfsub$, for
  substitutional transition metal solutes in fct afmD and bcc fm Fe.}
\end{table*}

\begin{table*}[htbp]
\begin{ruledtabular}
\begin{tabular}{lccccccc}
& $\ebind(\mathrm{vac,X;1a})$ & $\ebind(\mathrm{vac,X;1b})$ & $\ebind(\mathrm{vac,X;1c})$ & $\emig(\omega_2;\mathrm{1a})$ & $\emig(\omega_2;\mathrm{1b})$ & $\emig(\omega_1;\mathrm{1b}\rightarrow\mathrm{1b})$ & $\emig(\omega_1;\mathrm{1c}\rightarrow\mathrm{1c})$ \\ 
\hline
Sc & 0.750650 & 0.499434 & 0.505687 & 0.000000 & 0.004553 & 1.597425 & 1.203072 \\
Ti & 0.277210 & 0.106405 & 0.137513 & 0.036143 & 0.118099 & 1.150778 & 0.868661 \\
V & 0.095458 & -0.024377 & 0.002539 & 0.264122 & 0.421619 & 0.897122 & 0.717379 \\
Cr & 0.003838 & -0.074678 & -0.090630 & 0.559822 & 0.741722 & 0.731824 & 0.672854 \\
Mn & 0.004220 & -0.062333 & -0.069264 & 0.674926 & 1.097378 & 0.746450 & 0.740400 \\
Co & 0.023113 & 0.038358 & 0.010414 & 0.903159 & 1.142975 & 0.728135 & 0.685638 \\
Ni & 0.056450 & 0.027066 & 0.016082 & 0.891343 & 1.172443 & 0.779333 & 0.638315 \\
Cu & 0.121276 & 0.033691 & 0.071821 & 0.679113 & 0.939329 & 0.839959 & 1.084078 \\
Zn & 0.194651 & 0.063915 & 0.139639 & 0.469792 & 0.664598 & 0.909838 & 0.714105 \\
\hline
Y & 1.391114 & 1.146585 & 1.298059 & 0.000000 & 0.000000 & 2.225764 & 1.900111 \\
Zr & 0.885999 & 0.611256 & 0.624385 & 0.000000 & 0.000000 & 1.713237 & 1.354309 \\
Nb & 0.410330 & 0.178429 & 0.266261 & 0.039749 & --- & --- & --- \\
Mo & 0.224573 & 0.031336 & 0.071605 & 0.351614 & --- & --- & --- \\
Tc & 0.144515 & 0.013633 & -0.024236 & 0.700993 & --- & --- & --- \\
Ru & 0.137180 & 0.050221 & -0.021785 & 0.953080 & --- & --- & --- \\
Rh & 0.192630 & 0.096389 & 0.041094 & 1.007728 & --- & --- & --- \\
Pd & 0.276956 & 0.142563 & 0.140851 & 0.808536 & --- & --- & --- \\
Ag & 0.397634 & 0.207526 & 0.292348 & 0.500009 & --- & --- & --- \\
Cd & 0.514102 & 0.277870 & 0.413913 & 0.283986 & --- & --- & --- \\
\hline
Lu & 1.095221 & 0.816821 & 0.921025 & 0.000000 & 0.000000 & 1.929124 & 1.600727 \\
Hf & 0.688915 & 0.368512 & 0.455830 & 0.000000 & 0.000000 & 1.490638 & 1.214173 \\
Ta & 0.342817 & 0.115488 & 0.198550 & 0.129160 & --- & --- & --- \\
W & 0.190974 & 0.005138 & 0.036606 & 0.454051 & --- & --- & --- \\
Re & 0.120066 & -0.008993 & -0.047380 & 0.811482 & --- & --- & --- \\
Os & 0.121133 & 0.035760 & -0.056473 & 1.114752 & --- & --- & --- \\
Ir & 0.179482 & 0.095967 & 0.014834 & 1.216038 & --- & --- & --- \\
Pt & 0.277898 & 0.154282 & 0.126369 & 1.042552 & --- & --- & --- \\
Au & 0.418477 & 0.243583 & 0.291266 & 0.675957 & --- & --- & --- \\
Hg & 0.582178 & 0.353577 & 0.475884 & 0.349202 & --- & --- & --- \\
\end{tabular}
\end{ruledtabular}
\caption{\label{vacDataTab} Transition metal solute binding energies
  to a vacancy point defect, $\ebind(\mathrm{vac,X})$, in eV for the
  1a, 1b and 1c configurations, migration energies for vacancy-solute
  exchange, $\emig(\omega_2)$, in eV along paths 1a and 1b and
  migration energies for $\omega_1$ jumps, $\emig(\omega_1)$, in eV
  from configuration 1b to 1b and 1c to 1c [see
    \reffigs{DefectSubConfigFig}{fiveFreqFig}] in fct afmD Fe. A
  vacancy formation energy of 1.819197 eV was used to calculate the
  binding energies. Migration energies for $\omega_0$ jumps along
  paths 1a and 1b are 0.743409 and 1.048164 eV, respectively. For a Y
  solute, $\omega_1$ migration energies from configuration 1a to 1b
  and from 1b to 1a are 2.648928 and 2.404399 eV, respectively.}
\end{table*}

\begin{table*}[htbp]
\begin{ruledtabular}
\begin{tabular}{lcccccccc}
& $\ebind(\textrm{SI,X;mix-uu})$ & $\ebind(\textrm{SI,X;mix-ud})$ & $\ebind(\textrm{SI,X;1a})$ & $\ebind(\textrm{SI,X;1b})$ & $\ebind(\textrm{SI,X;1c})$ & $\ebind(\textrm{SI,X;2a})$ & $\ebind(\textrm{SI,X;2b})$ & $\ebind(\textrm{SI,X;2c})$ \\ 
\hline
Sc & unstable & unstable & 0.613195 & -0.148404 & -0.239088 & 0.082942 & 0.650971 & 0.233076 \\
Ti & -0.890140 & -1.001952 & 0.422515 & -0.026137 & -0.043963 & 0.029957 & 0.466075 & 0.114717 \\
V & -0.273120 & -0.385646 & 0.230051 & 0.114587 & 0.114224 & 0.001583 & 0.292839 & 0.031147 \\
Cr & 0.279108 & 0.196335 & 0.035055 & 0.188501 & 0.193289 & -0.036870 & 0.068725 & -0.004504 \\
Mn & 0.173584 & 0.149788 & 0.015833 & 0.032447 & 0.031454 & -0.035519 & 0.012747 & -0.001813 \\
Co & 0.218336 & -0.006592 & -0.042215 & 0.069918 & 0.035001 & -0.013967 & 0.003033 & 0.045561 \\
Ni & 0.064503 & -0.190004 & -0.006108 & 0.030123 & -0.077773 & -0.033833 & -0.011471 & 0.070368 \\
Cu & -0.322243 & -0.511532 & 0.037862 & -0.049706 & -0.168800 & -0.022084 & 0.050442 & 0.094807 \\
Zn & -0.565613 & -0.753655 & 0.095363 & -0.103980 & -0.196545 & -0.001840 & 0.155897 & 0.083301 \\
\hline
Y & unstable & unstable & 0.698495 & --- & --- & --- & 0.710074 & 0.315204 \\
Zr & unstable & unstable & 0.645110 & --- & --- & --- & 0.659849 & 0.222883 \\
Nb & -1.528735 & -1.599024 & 0.488008 & --- & --- & --- & 0.510657 & 0.123029 \\
Mo & -1.072197 & -1.187259 & 0.293452 & --- & --- & --- & 0.331194 & 0.064893 \\
Tc & -0.858870 & -0.952616 & 0.141055 & --- & --- & --- & 0.197384 & 0.054389 \\
Ru & -0.848137 & -0.863409 & 0.096056 & --- & --- & --- & 0.144930 & 0.080102 \\
Rh & -0.933354 & -0.950466 & 0.145075 & --- & --- & --- & 0.153587 & 0.150243 \\
Pd & -1.031908 & -1.105313 & 0.218741 & --- & --- & --- & 0.190158 & 0.192314 \\
Ag & unstable & -1.318686 & 0.303076 & --- & --- & --- & 0.308318 & 0.203754 \\
Cd & unstable & -1.556954 & 0.367269 & --- & --- & --- & 0.434429 & 0.186782 \\
\hline
Lu & unstable & unstable & 0.681841 & --- & --- & --- & 0.696565 & 0.271834 \\
Hf & unstable & unstable & 0.635332 & --- & --- & --- & 0.654131 & 0.212396 \\
Ta & -1.636926 & -1.715900 & 0.492322 & --- & --- & --- & 0.522191 & 0.119564 \\
W & -1.256014 & -1.389369 & 0.312929 & --- & --- & --- & 0.365106 & 0.054661 \\
Re & -0.914785 & -1.035476 & 0.132741 & --- & --- & --- & 0.217986 & 0.028394 \\
Os & -0.955385 & -1.031388 & 0.061411 & --- & --- & --- & 0.152669 & 0.042942 \\
Ir & -1.098984 & -1.064321 & 0.087307 & --- & --- & --- & 0.147562 & 0.102479 \\
Pt & -1.182566 & -1.200569 & 0.165787 & --- & --- & --- & 0.174715 & 0.155350 \\
Au & -1.358448 & -1.369936 & 0.253597 & --- & --- & --- & 0.262270 & 0.169659 \\
Hg & unstable & -1.555926 & 0.327290 & --- & --- & --- & 0.388016 & 0.154981 \\
\end{tabular}
\end{ruledtabular}
\caption{\label{siDataTab} Transition metal solute binding energies to
  an [001] self-interstitial dumbbell point defect,
  $\ebind(\mathrm{SI,X})$, in eV for the two distinct mixed dumbbells
  and for the 1a, 1b 1c, 2a, 2b and 2c configurations [see
    \reffig{DefectSubConfigFig}] in fct afmD Fe. For the mixed
  dumbbell the solute can lie between two up spin layers (mix-uu) or
  an up and down spin layer (mix-ud). A dumbbell formation energy of
  3.195402 eV was used to calculate the binding energies.}
\end{table*}

\begin{table}[htbp]
\begin{ruledtabular}
\begin{tabular}{cccc}
Site & $\ebind$ & Site & $\ebind$ \\
\hline
2a & -0.113799 & 3c & 0.168786 \\
2b & -0.152349 & 3d & 0.086246 \\
2c & -0.098449 & 4a & 0.218183 \\
3a &  0.005167 & 4b & 0.079361 \\
3b &  0.018938 & 4c & 0.106539 \\
\end{tabular}
\end{ruledtabular}
\caption{\label{EbYDataTab} Vacancy-Y binding energies, $\ebind$, in
  eV at 2nn, 3nn and 4nn separations [see \reffig{DefectSubConfigFig}]
  in fct afmD Fe.}
\end{table}

\begin{table}[htbp]
\begin{ruledtabular}
\begin{tabular}{cccc}
Jump & $\emig$ & Jump & $\emig$ \\
\hline
1a$\rightarrow$2a & 1.808723 & 1c$\rightarrow$3c & 1.208250 \\
1b$\rightarrow$2a & 2.134065 & 1c$\rightarrow$3d & 1.501190 \\
1c$\rightarrow$2c & 1.836946 & 1a$\rightarrow$4a & 1.261512 \\
1a$\rightarrow$3b & 1.659928 & 1c$\rightarrow$4c & 1.368114 \\
1b$\rightarrow$3b & 1.441606 & & \\
\end{tabular}
\end{ruledtabular}
\caption{\label{EmYDataTab} Migration energies, $\emig$, in eV for
  dissociative ($\omega_3$) vacancy jumps near a Y solute in fct afmD
  Fe. The jump paths are defined by the initial and final
  configurations [see
    \reffigs{DefectSubConfigFig}{fiveFreqFig}]. Migration energies for
  the reverse (dissociative, $\omega_4$) jumps can be calculated from
  these using the vacancy-Y binding energies in
  \reftabs{vacDataTab}{EbYDataTab}. Jumps where the migrating Fe atom
  would be constrained (by the collinear calculations) to have zero
  moment at some point on the path have not been calculated as they
  would result in a significant overestimation of the migration
  energy\cite{Klaver12}.}
\end{table}

\end{document}